\documentclass[]{aastex7}

\usepackage{amsmath,amssymb}
\usepackage{booktabs}
\begin{document}

\setcitestyle{square}

\title{Planar Collisionless Shock Simulations with Semi-Implicit Particle-in-Cell Model FLEKS}

\author[orcid=0000-0003-4571-4501,sname=Zhou]{Hongyang Zhou}
%\altaffiliation{Center for Space Physics}
\affiliation{Center for Space Physics, Boston University, Boston, MA 02215, USA}
\email[show]{hongyang@bu.edu}  

\author[sname=Chen]{Yuxi Chen}
\affiliation{Center for Space Environment Modeling, University of Michigan, Ann Arbor, MI 48109, USA}
\email{yuxichen@umich.edu}

\author[orcid=0000-0002-8990-094X,sname=Dong]{Chuanfei Dong}
\affiliation{Center for Space Physics, Boston University, Boston, MA 02215, USA}
\email{dcfy@bu.edu}

\author[orcid=0000-0003-4821-3120,sname=Wang]{Liang Wang}
\affiliation{Center for Space Physics, Boston University, Boston, MA 02215, USA}
\email{wang0734@bu.edu}

\author[sname=Zou]{Ying Zou}
\affiliation{Johns Hopkins University Applied Physics Laboratory, Laurel, MD 20723, USA}
\email{Ying.Zou@jhuapl.edu}

\author[sname=Walsh]{Brian M. Walsh}
\affiliation{Center for Space Physics, Boston University, Boston, MA 02215, USA}
\affiliation{Department of Mechanical Engineering, Boston University, Boston, MA 02215, USA}
\email{bwalsh@bu.edu}

\author[sname=Tóth]{Gábor Tóth}
\affiliation{Center for Space Environment Modeling, University of Michigan, Ann Arbor, MI 48109, USA}
\email{gtoth@umich.edu}

\begin{abstract}

This study investigates the applicability of the semi-implicit particle-in-cell code \texttt{FLEKS} to heliospheric shock simulations.  We examine one- and two-dimensional local planar shock simulations, initialized using MHD states with upstream conditions representative of plasmas in the hypersonic, $\beta\sim 1$ regime, for both quasi-perpendicular and quasi-parallel configurations.  The refined algorithm in \texttt{FLEKS} proves robust, enabling accurate shock simulations with a grid resolution on the order of the electron inertial length $d_e$.  Our simulations successfully capture key shock features, including shock structures (foot, ramp, overshoot, and undershoot), upstream and downstream waves (fast magnetosonic, whistler, Alfvén ion-cyclotron, and mirror modes), and non-Maxwellian particle distributions.  Crucially, we find that at least two spatial dimensions are critical for accurately reproducing downstream wave physics in quasi-perpendicular shocks and capturing the complex dynamics of quasi-parallel shocks, including surface rippling, shocklets, SLAMS, magnetic reconnection and jets.  Furthermore, our parameter studies demonstrate the impact of mass ratio and grid resolution on shock physics.  This work provides valuable guidance for selecting appropriate physical and numerical parameters for shock simulations using a semi-implicit PIC method, paving the way for incorporating kinetic shock processes into large-scale collisionless plasma simulations with the MHD-AEPIC model.

\end{abstract}

\keywords{\uat{Space plasmas}{1544} --- \uat{Plasma astrophysics}{1261} --- \uat{Shocks}{2086}}


\section{Introduction} 

Collisionless shocks are ubiquitous in space plasmas, marking abrupt transitions in the velocity of charged particles.
Unlike shocks in neutral fluids, collisionless shocks are mediated by electromagnetic fields rather than particle collisions.
These shocks play a crucial role in diverse astrophysical environments, including astrophysical jets \citep{livio1999astrophysical, blandford2019relativistic}, the solar wind \citep{oliveira2023geoeffectiveness}, the heliosphere \citep{stone1985collisionless}, and planetary magnetospheres (see more in \citet{balogh2013physics} and references therein).
The passage of a shock can dramatically alter the properties of a plasma, leading to particle acceleration, energy dissipation, and the generation of various wave modes across a broad spectrum of frequencies.

While magnetohydrodynamics (MHD) provides a fluid description of shocks, it treats them as simple discontinuities and fails to capture the essential kinetic processes that govern their microscopic structure and evolution.
To accurately model collisionless shocks, kinetic approaches are necessary, such as hybrid simulations \citep{winske2003hybrid}, which treat ions kinetically and electrons as a fluid, or full particle-in-cell (PIC) simulations \citep{birdsall2018plasma}, which treat both ions and electrons as discrete particles.
Particle-based methods, while powerful, face challenges related to statistical noise, artificial heating/cooling, and numerical instabilities, especially when simulating large-scale systems with vast plasma scale separation.

Recent developments in PIC methods have led to semi-implicit algorithms that offer a promising pathway to overcome some of these long-standing limitations. These methods relax the stringent stability constraints of explicit PIC while avoiding the computationally expensive nonlinear iterations required in fully implicit PIC \citep{lapenta2023advance, ren2024recent}. Semi-implicit PIC methods, by allowing for larger grid cells and time steps, offer improved energy conservation and numerical stability, making them more amenable to simulating the larger physical domains characteristic of astrophysical scenarios. This study investigates the applicability of a semi-implicit, energy and charge conserving PIC model, \texttt{FLEKS} \citep{chen2023fleks}, to collisionless shock simulations. \texttt{FLEKS} offers several key advantages, including near-exact energy and charge conservation \citep{lapenta2017exactly, chen2019gauss} and the ability to use significantly larger grid sizes and time steps compared to traditional explicit PIC methods.
However, for shock tests with the original semi-implicit algorithm, we found that the super-Alfvénic upstream region shows unstable wave growth non-existing in nature. These so-called "spurious" oscillations are essentially associated with particle noise that gets amplified in the solution of the electric field from the semi-implicit solver.

We have proposed to address numerical instabilities encountered in shock simulations by (a) adding a Lax-Friedrichs-type diffusion term to the electromagnetic field solvers to suppress spurious oscillations near discontinuities and (b) using a novel particle current calculation method to reduce statistical noise \citep{chen2025fleks}.
In this study, we present one-dimensional (1D) and two-dimensional (2D) local planar shock simulations using the refined \texttt{FLEKS} algorithm within the Space Weather Modeling Framework (SWMF, \citet{toth2012adaptive}).
The simulations focus on non-relativistic, heliospheric shocks, typically characterized by Alfvén Mach numbers $\mathrm{M}_\mathrm{A} \lesssim \mathcal{O}(10)$. Astrophysical shocks, such as those associated with supernova remnants, can exhibit much higher Mach numbers, often reaching tens or hundreds. In such high-$\mathrm{M}_\mathrm{A}$ regimes, different kinetic instabilities, like the Buneman instability \citep{matsumoto2012electron}, Weibel instability \citep{kropotina2023weibel}, and Bell instability \citep{park2015simultaneous}, may become dominant in structuring the shock front and adjacent regions. This sensitivity of plasma physics to the specific parameter regime underscores the importance of conducting targeted simulations to investigate phenomena within distinct environments.
For heliospheric shocks, we explore both quasi-perpendicular and quasi-parallel configurations, varying the ion-to-electron mass ratio $m_i/m_e$ and grid resolution $\Delta x$ on the order of the electron skin depth $d_e$.
Our results demonstrate that the refined \texttt{FLEKS} algorithm accurately reproduces key hypersonic shock features with conditions representative at 1 AU, including the detailed shock structures, the generation of various wave modes (e.g., fast magnetosonic, whistler, Alfvén ion-cyclotron, and mirror modes), and the formation of non-Maxwellian particle distributions, provided that the resolution is on a sub-ion scale and a 2D spatial domain is used.
These findings validate the enhanced \texttt{FLEKS} capabilities for collisionless shock simulations and offer crucial guidance for model parameter selection in future studies.
Ultimately, this work represents a significant step towards incorporating detailed kinetic shock physics into large-scale, global plasma models, with direct implications for understanding particle acceleration and energy conversion in diverse astrophysical environments beyond just terrestrial applications.

\section{Methodology} \label{sec:method}

%(Full equations shall be shown in the algorithm paper.) To ensure numerical stability, we solve the implicit electric field equation in the comoving frame, separating the motional electric field $\mathbf{E}_m = -\mathbf{U}\times\mathbf{B}$ from the other contributions, where $\mathbf{U}$ and $\mathbf{B}$ are the local bulk velocity and magnetic field. In addition, we introduce a Lax–Friedrichs type scheme for the Maxwell's equations, described in detail in \cite{chen2025fleks}, with parameters $\theta = 0$ and $\text{isotropy} = 1$. Note that the isotropy parameter has no effect in 1D simulations.

The \textbf{FL}exible \textbf{E}xascale \textbf{K}inetic \textbf{S}imulator (\texttt{FLEKS}) is a semi-implicit PIC code that utilizes a Gauss's Law-satisfying Energy-Conserving Semi-Implicit Method (GL-ECSIM).
In \texttt{FLEKS}, Maxwell's equations are solved on a collocated staggered grid, where electric fields are at grid nodes and magnetic fields at cell centers.
To address spurious oscillations near discontinuities and statistical noise in fast plasma flows, recent \texttt{FLEKS} version incorporates two key numerical techniques.
First, a Lax-Friedrichs-type diffusion term with a flux limiter is introduced into the Maxwell solver to suppress unphysical oscillations.
Second, a novel approach calculates the current density in the comoving frame, which significantly reduces particle noise in simulations with fast plasma flows.
Digital filter smoothing is also applied to the comoving frame current density to further enhance numerical stability and mitigate numerical cooling effects.
Readers interested in more details are referred to \cite{chen2025fleks}.

Collisionless shocks in space are characterized by several dimensionless parameters, including the Alfvén Mach number ($\text{M}_\text{A}$) and the electron and ion plasma betas ($\beta_e$ and $\beta_i$).
These parameters govern the fundamental properties of shocks and can vary over orders of magnitude across all kinds of astrophysical shocks.
This study primarily investigates upstream plasma conditions typical of the solar wind at 1 AU near the terrestrial magnetosphere. We selected an average plasma state with $\text{M}_\text{A} = 7$,\,$\beta_i = 0.5$, and $\beta_e = 1.0$, based on solar wind temperature statistics from NASA's Wind spacecraft as reported by \citet{salem2023precision} and summarized in their Figure 9 for the dependence on solar wind speed and $\beta$.
Note that similar parameters in the hypersonic plasma regime can occur at other solar system planets (e.g. Mars \citep{liu2021statistical}) and also astrophysical systems, which presents a broader scope outside Earth.
%A complete exploration of the vast upstream plasma conditions is out of the scope of the current study.

To reduce computational cost, we employed an artificially reduced proton-to-electron mass ratio of $m_i / m_e = 25$ and a ratio of electron plasma frequency to electron cyclotron frequency of $\omega_{pe}^2 / \omega_{ce}^2 = 10$.
Since $\omega_{pe}/\omega_{ce} = c/V_{Ae}$, where $V_{Ae} = B / \sqrt{\mu_0 \rho_e} = V_{Ai}\sqrt{m_i / m_e}$ is the electron Alfvén speed, these ratios determine the effective speed of light and other key plasma scales in the simulation.
With an upstream plasma number density of $n = n_i = n_e = 3\,\text{cm}^{-3}$ and magnetic field of $B = 5\,\text{nT}$, the system is fully determined.
Table \ref{tbl-1} compares the electron and ion spatiotemporal scales in the simulation to their corresponding values in nature.
Here, $\lambda_D$ is the Debye length, $d_i$ is the ion inertial length, $r_{ci}$ is the ion gyroradius, $d_e$ is the electron inertial length, $r_{ce}$ is the electron gyroradius, $\omega_{pe}$ is the electron gyrofrequency, $\omega_{ci}$ is the ion cyclotron frequency, and $\omega_{pi}$ is the ion gyrofrequency.
In the following context, all the characteristic quantities are referred to the upstream values ("up") unless otherwise specified.

% $T_e = 2.4\times 10^5\,\text{K}$, $T_i = 1.2\times 10^5\,\text{K}$
\begin{table}
\caption{Plasma characteristic scales upstream of the 1D and 2D simulations assuming $m_i / m_e = 25$, $\beta_i = 0.5$, $\beta_e = 1$, $n_i = n_e = 3\,\text{cm}^{-3}$, $B=5\,\text{nT}$. The simulated speed of light is set to $c=3\times 10^6\,\text{m/s}$, which is 1/100 the vacuum speed of light.
}
\label{tbl-1}
\centering
\begin{tabular}{c c c}
\hline
 Upstream Parameter  & Real Value & Simulation Value \\
\hline
  $\lambda_D$  & 0.02 km  & 5.9 km   \\
  $d_i$  & 133 km  & 133 km   \\
  $r_{ci}$  & 114 km  & 114 km   \\
  $d_e$  & 3.1 km  & 26 km   \\
  $r_{ce}$  & 2.6 km  & 23 km   \\
  $2\pi/\omega_{pe}$ & $6.4\times 10^{-5}$ s  & $0.17$ s  \\
  $2\pi/\omega_{ci}$ & $13$ s & $13$ s  \\
  $2\pi/\omega_{ce}$ & $7\times 10^{-3}$ s & $0.52$ s  \\
\hline
%\multicolumn{2}{l}{$^{a}$Footnote text here.}
\end{tabular}
\end{table}

\citet{lembege2003full} provides a comprehensive summary of shock initialization methods, including the magnetic piston method \citep{lembege1987self}, the injection method (e.g., \citet{leroy1982shock, burgess1989shock, quest1985shock, ofman2009shock}), the flow-flow method \citep{omidi1992kinetic}, and the plasma release method \citep{ohsawa1985strong}.
While the injection method, which generates a shock by launching a particle beam against a reflecting boundary, is widely used in local shock simulations, our approach aligns more closely with the relaxation method, also known as the shock-rest-frame model, commonly employed in hybrid \citep{lee2017generation, pfaukempf2018vlasov} and full-PIC \citep{umeda2006full} simulations.
In this method, we first determine the initial downstream MHD plasma parameters and magnetic field using the upstream conditions and the Rankine-Hugoniot (RH) jump conditions.
We then solve the ideal MHD equations, augmented with a scalar electron pressure equation (with an adiabatic index $\gamma=5/3$), using the \texttt{BATS-R-US} MHD model \citep{toth2012adaptive} until a steady-state solution is achieved.
This MHD solution is then interpolated onto the PIC grid, assuming isotropic Maxwellian particle distributions, coupled through SWMF.
Finally, the system is evolved by solving the Vlasov-Maxwell equations for electrons and ions in the PIC domain using \texttt{FLEKS}.
In the following analysis, the displayed time refers only to the PIC simulation phase.

Table \ref{tbl-2} lists the planar shock simulations performed in this study.
The electron skin depth is related to the ion inertial length as $d_e = \sqrt{m_e/m_i}\,d_i$.
As a convention, we define the +x direction (right) as upstream and the -x direction (left) as downstream, with the inflow plasma moving from right to left.
The shock-normal angle $\theta_{Bn,\text{up}}$ defines the orientation between the upstream magnetic field and the shock normal. It is given by $\theta_{Bn,\text{up}} = \tan^{-1}B_{t,\text{up}} / B_{n,\text{up}}$, where $B_{t,\text{up}}$ and $B_{n,\text{up}}$ are the tangential and normal components of the upstream magnetic field with respect to the shock.
A fixed Maxwellian boundary condition is applied at the upstream inflow face (+x), while a floating boundary condition is used at the downstream face (-x).
Periodic boundary conditions are applied for all other directions.

\begin{table}
\caption{List of planar shock runs. $\theta_{Bn,\text{up}}$ is the upstream shock-normal angle, $\mathrm{M}_\mathrm{A,up}$ is the upstream Mach number, $m_i / m_e$ is the mass ratio, and $\Delta x$ is the grid resolution normalized by the upstream electron inertial length.}
\label{tbl-2}
\centering
\begin{tabular}{c c c c c c}
\toprule
 Run  & Dimension & $\theta_{Bn,\text{up}}\,[^\circ]$ & $\mathrm{M}_\mathrm{A,up}$ & $m_i / m_e$ & $\Delta x\,[d_{e,\text{up}}]$  \\
\midrule
1  & 1 & 87 & 7 & 25  & 1 \\
2  & 1 & 30 & 7 & 25  & 1 \\
3  & 1 & 87 & 7 & 100 & 1 \\
4  & 1 & 87 & 7 & 400 & 1 \\
5  & 1 & 30 & 7 & 100 & 1 \\
6  & 1 & 30 & 7 & 400 & 1 \\
7  & 1 & 87 & 7 & 25  & 0.5 \\
8  & 1 & 87 & 7 & 25  & 2 \\
9  & 1 & 87 & 7 & 25  & 4 \\
10 & 1 & 30 & 7 & 25  & 0.5 \\
11 & 1 & 30 & 7 & 25  & 2 \\
12 & 1 & 30 & 7 & 25  & 4 \\
13 & 2 & 87 & 7 & 25  & 1 \\
14 & 2 & 30 & 7 & 25  & 1 \\
15 & 2 & 30 & 12  & 25 & 1 \\
16 & 2 & 0 & 7 & 25 & 1 \\
\bottomrule
%\multicolumn{2}{l}{$^{a}$Footnote text here.}
\end{tabular}
\end{table}

Unlike global magnetosphere simulations, which often use the planetary radius as a characteristic length scale, local shock simulations lack an inherent spatial reference.
Therefore, we typically normalize quantities to the relevant kinetic scales of ions and electrons.
To emphasize the scaling relations in the local system, all displayed plasma quantities are normalized to characteristic upstream ion scales: $d_{i,\text{up}}$ for length, $\omega_{ci,\text{up}}$ for frequency, $n_{i,\text{up}}$ for density, $V_{Ai,\text{up}}$ for velocity, and $P_{i,\text{up}}$ for pressure, unless otherwise specified.
Electromagnetic fields are normalized to the magnitudes of the upstream values $B_{\text{up}}$ and $E_{\text{up}}=|\mathbf{V}_{Ai,\text{up}} \times \mathbf{B}_{\text{up}}|$.
All the displayed times are normalized by the upstream proton gyrofrequency $\omega_{ci,\text{up}}$, where a $2\pi$ factor is ignored for converting from gyrofrequencies to periods.
The output cadence is set to $1\,\text{s} \simeq 0.08\,\omega_{ci,\text{up}}^{-1}$.

As a PIC model, \texttt{FLEKS} unavoidably exhibits statistical noise due to the use of a finite number of macro-particles that scales as $\propto N^{-1/2}$, where $N$ is the number of macro-particles per cell.
Empirically we find a noise level $\sim 10\%$ for $N=100$ and $\sim 5\%$ for $N=400$ from a 1D free-stream test with uniform plasma states, electromagnetic fields, and periodic boundary conditions.
In the 1D and 2D shock simulations presented here, we used $N=400$, a value that remains feasible for global 2D magnetosphere simulations while effectively suppressing noise.
While techniques like high-order shape functions \citep{muralikrishnan2021sparse} or kernel density estimation \citep{wu2018reducing} can help mitigate noise, they introduce additional complexities in performance and implementation.
Therefore, we currently employ the standard second-order scheme with a triangular shape function.
To avoid visual bias in velocity space plots, we applied an ion cutoff density value of $w_\text{min}=10^{-3}w_\text{peak}$ for quasi-perpendicular shock simulations and $w_\text{min}=10^{-4}w_\text{peak}$ for quasi-parallel shock simulations, where $w_\text{peak}$ is the maximum phase space density in the sampled region.
Phase space densities below these thresholds were set to 0 and displayed as white in the logarithmic scale colormaps.
A 3-point sliding window smoothing is applied for showing the plasma moments in the line profiles.
%We also tested the performance of kernel density estimation on the phase space distributions from particle sampling.
%It turns out that using the default kernel selection method in $\textit{scipy gaussian kde}$ tends to overestimate the temperature, especially in the presence of beam distributions.

\section{Results} \label{sec:result}

\subsection{Local 1D Simulations}

We start the investigation with simple 1D geometries.
For both quasi-perpendicular and quasi-parallel shock cases, we typically employed a PIC grid resolution of $\Delta x = 26\,\text{km}$, which corresponds to 1 upstream electron inertial length $d_{e,\text{up}}$ or $1/5$ of the upstream ion inertial length $d_{i,\text{up}}$.
This resolution adequately captures the ion-scale dynamics while also accounting for
electron-scale effects.
To accommodate the development of upstream foreshock waves in the quasi-parallel simulations, we used a larger domain extent of $400\,d_{i,\text{up}}$ in the shock-normal direction, compared to $200\,d_{i,\text{up}}$ for the quasi-perpendicular cases. Each \texttt{FLEKS} run lasts for $20\,\omega_{ci,\text{up}}^{-1}=262\,\text{s}$.

\subsubsection{1D Quasi-Perpendicular Shocks} \label{ssec:1dqperp}

\begin{figure}
\noindent\includegraphics[width=\textwidth]{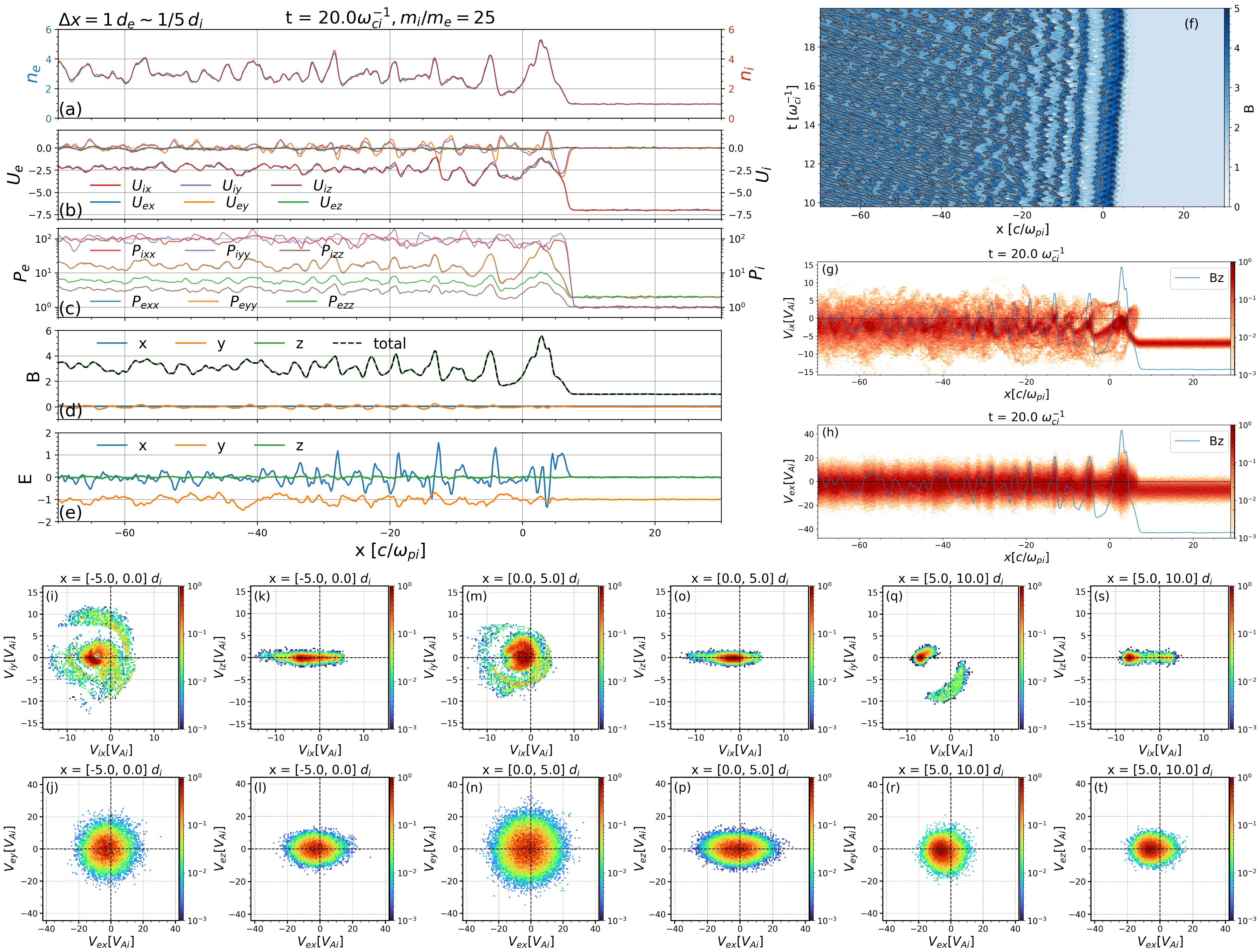}
\caption{1D simulation of a quasi-perpendicular shock with $\theta_{Bn} = 87^\circ$ and $m_i / m_e = 25$. The plasma moments and EM fields are normalized by the upstream quantities. Phase space distributions are normalized by the maximum phase space density in the displayed regions. (a-e) Close-up of plasma quantities and electromagnetic fields near the shock front at $t=20\,\omega_{ci,\text{up}}$. (f) $x-t$ stack plot of the total magnetic field, overlaid with ion density contours. (g,h) Ion and electron $x-v_{x}$ phase space plot, with the $B_z$ profile shown in blue for reference. Ion and electron velocity space distributions are displayed in the downstream region (i-l), downstream right behind the shock front (m-p), and upstream right next to the shock front (q-t).}
\label{fig1}
\end{figure}

Figure \ref{fig1} presents a detailed view of plasma moments, electromagnetic fields, and electron and ion phase space distributions near the quasi-perpendicular $\theta_{Bn} = 87^\circ$ shock at $t=20\,\omega_{ci,\text{up}}$, with the shock front located around $x=5\,d_{i,\text{up}}$.
The characteristic shock structures, including the foot, ramp, overshoot, and undershoot, are clearly discernible in the density (Figure \ref{fig1}a) and magnetic field (Figure \ref{fig1}d) profiles.
The ramp width, defined as the transition from the onset of magnetic field variation to the first maximum (overshoot), measures approximately 0.5 ion convective gyroradii $r_{ci,\text{up}} = V_{\text{up}} / \omega_{ci,\text{up}}$, or 4 upstream ion inertial lengths $d_{i,\text{up}} = c / \omega_{pi,\text{up}}$. 
Due to shock drift acceleration by the upstream convection electric field (primarily $E_y$ in Figure \ref{fig1}e), reflected protons exhibit higher energy than directly transmitted protons.
Their gyration downstream results in arches and holes in the $x-V_{ix}$ phase space (Figure \ref{fig1}g) at large $V_{ix}$ magnitudes, although these features are partially smoothed by the thermal velocity.
The cross-shock electric potential, arising from charge separation due to the abrupt magnetic field jump, can be estimated from the electric field $E_x$ in Figure \ref{fig1}e as $\Delta \phi = \int E_x \mathrm{d}x \simeq 430\,\text{V}$.
This yields an electric potential energy $U_E = q_i\Delta \phi \approx 0.4 E_k$, where $E_k=m_i V_{i,\text{up}}^2/2$ is the upstream ion kinetic energy.

The preshock region in this quasi-perpendicular shock is notably quiescent, as the Alfvén Mach number $\text{M}_\text{A}=7$ significantly exceeds the whistler critical Mach number $\text{M}_\text{w}$ \citep{kennel1985quarter, krasnoselskikh2010theory}, defined by
\begin{linenomath*}
\begin{equation} \label{eq-whistler-mach}
\text{M}_\text{w} \equiv \frac{1}{2}|\cos\theta_{Bn}|\sqrt{\frac{m_i}{m_e}}
\end{equation}
\end{linenomath*}
Consequently, whistler precursors are unable to form and propagate upstream in this simulation.

Although the initial MHD conditions for our PIC simulations satisfy the stationarity requirement, i.e., pressure balance between magnetic, ram, and thermal pressures ($P_B + P_\text{ram} + P_\text{th} = \text{const.}$) across the shock front, this balance is disrupted as the shock evolves under the fully kinetic solver in the supercritical regime ($M_A > M_c \sim 2.5$, where $M_c$ is the first critical Mach number described in \citet{leroy1982shock}).
In this simulation, the shock front moves less than $10\,d_{i,\text{up}} \simeq 1320\,\text{km}$ in $20\,\omega_{ci,\text{up}}^{-1} \simeq 262\,\text{s}$ in the lab frame, corresponding to a speed below $5\,\text{km/s}$ along x, indicating that the perpendicular shock remains quasi-steady.
The time evolution of the total magnetic field strength from $t=10\,\omega_{ci,\text{up}}^{-1}$ to $20\,\omega_{ci,\text{up}}^{-1}$ is depicted in the stack plot (Figure \ref{fig1}f), overlaid with iso-density lines.
The overshoots (dark blue) and undershoots (white) are clearly visible across the shock front.
Particularly, magnetic and density oscillations within $20\,d_{i,\text{up}}$ near downstream move synchronously with the shock front, suggesting they are stationary structures rather than propagating waves.
Further downstream, the slope of the iso-density contours coincides with the plasma bulk velocity, indicating their origin as the structures moving with the plasma flow.

Correlated periodic downstream fluctuations are observed in the plasma density, total magnetic field, and electron thermal pressures (Figure \ref{fig1}a, \ref{fig1}c, and \ref{fig1}d).
Ion gyration and the formation of downstream ring distributions (Figure \ref{fig1}i and \ref{fig1}m) are intrinsic consequences of ion dynamics across the shock front.
In the downstream region, most ions directly penetrate the shock front and are decelerated by the cross-shock potential.
The fraction of reflected ions is small (Figure \ref{fig1}g), limited to a normal length of $2\,d_{i,\text{up}}$ in the ramp.
Minimal ion deflection by the magnetic field occurs due to the ramp width being much smaller than the upstream ion convective gyroradius $r_{ci,\text{up}}=V_{i,\text{up}}/\omega_{ci,\text{up}}$.
These non-isotropic reflected ions (Figure \ref{fig1}g, \ref{fig1}q-r), after being drift-accelerated along the y-direction, recross the ramp and form a downstream gyrating beam, leading to oscillating ion pressure.
The spatial scale of the periodic structure $L$ corresponds to the ion ring distribution in Figure \ref{fig1}i, centered at $V_{ix}=-2V_{Ai,\text{up}}, V_{iy}=0$ with a radius of $V_{ring} \simeq 8V_{Ai,\text{up}}$, where $L \sim 2.4\,d_{i,\text{up}}\sim m_i V_{ring}/ (q_i B_{\text{down}})$).
Due to their frozen-in behavior with the background magnetic field, the total magnetic field also exhibits coherent periodicity (Figure \ref{fig1}d).
With suppressed wave-particle interactions due to the 1D configuration and slow gyrophase mixing, this postshock periodic structure persists nearly to the downstream simulation boundary.

We conducted a companion simulation with $\Delta x = 0.5\,d_e$ and confirmed that the downstream periodic structures are ion-scale phenomena, independent of electron-scale numerical convergence.
However, the inclusion of electron kinetics in \texttt{FLEKS} introduces electrostatic electron pressure modulation by ion and magnetic pressure variations, evident in the correlated spatial oscillation of the $V_{ex}$ extent (Figure \ref{fig1}h) and the modulated pressure components $P_{exx}$, $P_{eyy}$, and $P_{ezz}$ (Figure \ref{fig1}c and \ref{fig1}h).

Three locations, centered downstream at $x=-2.5\,d_{i,\text{up}}$ (Figure \ref{fig1}i-l), across the shock at $x=2.5\,d_{i,\text{up}}$ (Figure \ref{fig1}m-p), and upstream in front of the shock at $x=7.5\,d_{i,\text{up}}$ (Figure \ref{fig1}q-t), each spanning $5\,d_{i,\text{up}}$, are selected to examine the velocity space distributions.
The three-dimensional velocity space distributions are projected onto the $V_x-V_y$ and $V_x-V_z$ planes, where the z-axis is approximately aligned with the background magnetic field.
The 1D simulation neglects parallel heating along the z-direction.
Consequently, no proton thermalization occurs along $V_{iz}$ in the downstream and ramp regions (Figure \ref{fig1}j and \ref{fig1}n).
This leads to highly anisotropic downstream ion distributions with $P_{i\perp}/P_{i\parallel} > 8$.
Electron distributions exhibit a similar lack of parallel heating in $V_{ez}$ (Figure \ref{fig1}l and p), but rapid thermalization occurs in the perpendicular directions, $V_{ex}$ and $V_{ey}$, immediately downstream of the ramp (Figure \ref{fig1}c, \ref{fig1}k, \ref{fig1}o, and \ref{fig1}s).

% It is important to recognize that this 1D simulation neglects parallel heating along the z-direction.
% Consequently, no proton thermalization occurs along $V_{iz}$ in the downstream and ramp regions (Figure \ref{fig1}j and \ref{fig1}n).
% This leads to highly anisotropic downstream ion distributions with $P_{i\perp}/P_{i\parallel} > 8$, an artifact of the reduced dimensionality.
% Electron distributions exhibit a similar lack of parallel heating in $V_{ez}$ (Figure \ref{fig1}l and p), but rapid thermalization occurs in the perpendicular directions, $V_{ex}$ and $V_{ey}$, immediately downstream of the ramp (Figure \ref{fig1}c, \ref{fig1}k, \ref{fig1}o, and \ref{fig1}s).

% As demonstrated later in the 2D quasi-perpendicular simulation (Figure \ref{fig3}), downstream parallel and oblique propagating waves, absent in the 1D case, play a crucial role in redistributing free energy associated with temperature anisotropy.
% A detailed discussion of these kinetic waves in the downstream region of quasi-perpendicular shocks is presented in Section \ref{sssec:2dqperp}.

\subsubsection{1D Quasi-Parallel Shocks}\label{sssec:1dqpar}

\begin{figure}
\noindent\includegraphics[width=\textwidth]{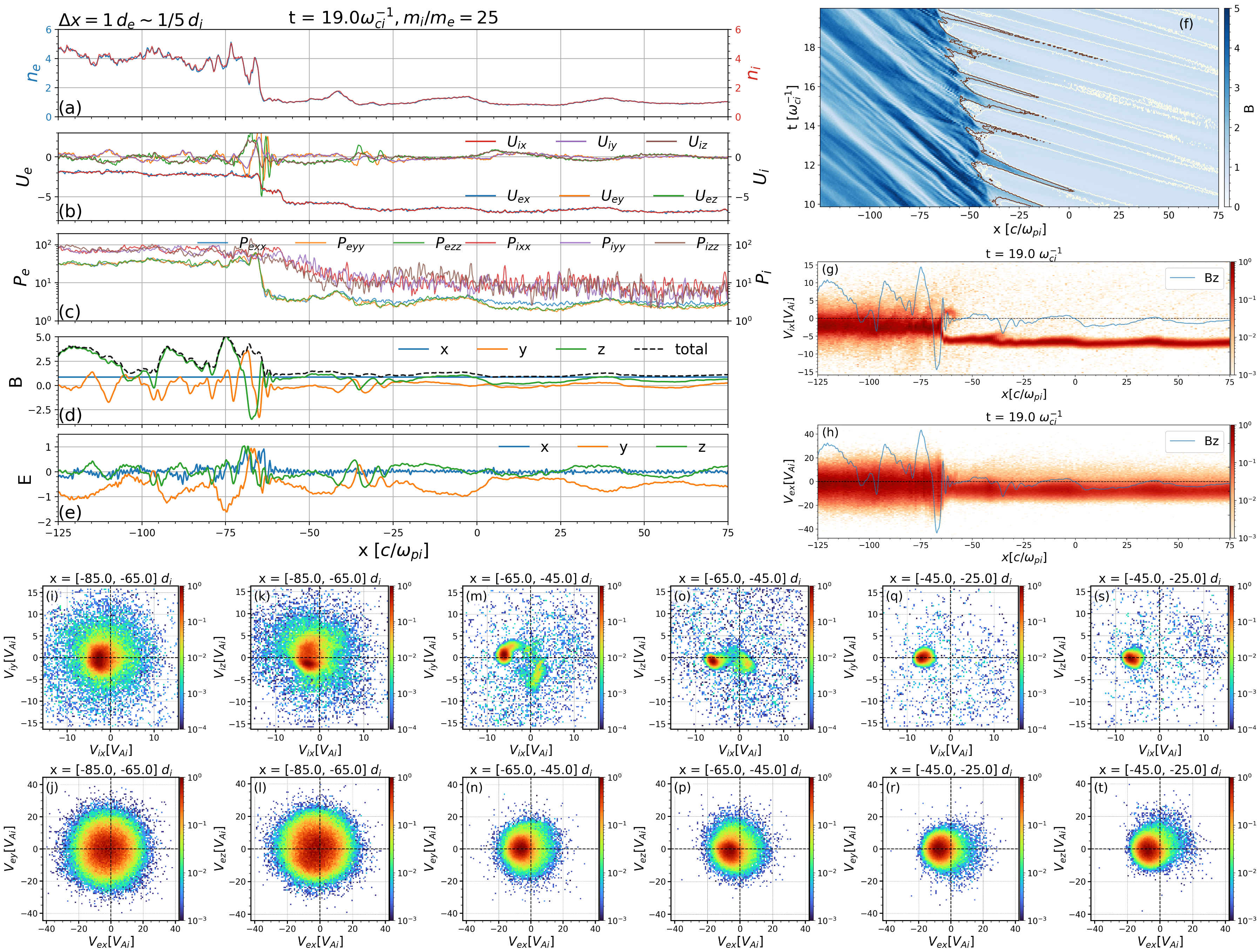}
\caption{1D simulation of a quasi-parallel shock with $\theta_{Bn} = 30^\circ$ and $m_i / m_e = 25$. The plasma moments and EM fields are normalized by the upstream quantities. Phase space distributions are normalized by the maximum phase space density in the displayed regions. (a-e) Close-up of plasma quantities and electromagnetic fields near the shock front at $t=19\,\omega_{ci,\text{up}}^{-1}$. (f) $x-t$ stack plot of the total magnetic field, overlaid with ion density contours. (g,h) Ion and electron $x-v_{x}$ phase space plot, with the $B_z$ profile shown in blue for reference. (i-l) Downstream ion and electron velocity space distributions. (m-p) Ion and electron velocity space distributions at the shock front. (q-t) Upstream ion and electron velocity space distributions.}
\label{fig2}
\end{figure}

In the 1D quasi-parallel shock run, we set a wider x extent from $-150\,d_{i,\text{up}}$ to $250\,d_{i,\text{up}}$ compared to the quasi-perpendicular case.
This guarantees enough space for the evolution of reflected ions at the shock as well as the growth of upstream waves.
Figure \ref{fig2} illustrates the results of a 1D quasi-parallel shock simulation with $\theta_{Bn} = 30^\circ$, presented in the same format as Figure \ref{fig1}.
Notably, this simulation reveals significantly more dynamic wave activity than the quasi-perpendicular case, evident in both the plasma quantities (Figure \ref{fig2}a-c) and electromagnetic fields (Figure \ref{fig2}d-e).
Within the upstream region ($x > -60\,d_{i,\text{up}}$), a dominant large-scale linearly polarized ultra-low frequency (ULF) wave extends far upstream and triggers perturbations in all variables.
It possesses significant in-phase density and magnetic field perturbations (Figure \ref{fig2}a and \ref{fig2}d).
Analysis of its phase velocity reveals that this wave mode propagate slightly slower than the inflow velocity in the lab frame.
Transforming to the plasma rest frame (i.e., the frame moving with the inflow velocity), the phase velocities reverse direction, now pointing upstream in the +x direction.
We thus identify this large-scale wave as the foreshock fast magnetosonic mode (e.g. \citet{eastwood2005foreshock}).
Corresponding to these wave fluctuations, Figure \ref{fig2}f, a spatiotemporal $x-t$ magnetic field stack plot, demonstrates the cyclic reformation of the main shock front from $t = 10\,\omega_{ci,\text{up}}^{-1}$ to $20\,\omega_{ci,\text{up}}^{-1}$.
This reformation is an unsteady process, driven by the nonlinear development of upstream waves.
The match of the extension of upstream enhanced magnetic field with the density peaks, shown by the iso-density colored lines, again confirms its magnetosonic property.
Furthermore, the downstream magnetic fluctuations exhibit higher amplitudes and broader spatial extents, consistent with observational data from Earth's quasi-parallel magnetosheath (e.g., \citet{lucek2002parallel}).

Compared to the quasi-perpendicular shock, the ion distribution in the quasi-parallel shock exhibits a more diffusive structure, a direct consequence of wave-particle interactions linked to foreshock wave generation.
The $x-V_{ix}$ phase plot (Figure \ref{fig2}g) illustrates reflected ions, characterized by a beam extending approximately $5\,d_{i,\text{up}}$ in the normal x-direction with $\lesssim 10\%$ of the inflow ion beam density (between $x \in [-65, -60]\,d_{i,\text{up}}$), alongside a wider distribution of diffuse populations in the upstream region.
The beam structure changes with time rapidly.
The fluctuations in the ion and electron phase space are also associated with the upstream ULF waves (Figure \ref{fig2}g-h, $x \in [-60, 75]\,d_{i,\text{up}}$), suggesting particle distribution modulation through Landau or cyclotron resonance via the parallel electric field ($ E_\parallel$ inferred from $E_x$ and $E_y$ in Figure \ref{fig2}e).

Downstream ions exhibit a nearly isotropic distribution (Figure \ref{fig2}i-j), with the core upstream population being heated by an order of magnitude across the shock (Figure \ref{fig2}c).
In contrast, upstream ions (Figure \ref{fig2}q-r) show a sparse diffuse distribution alongside the core inflow population.
Note that the colored ion velocity space densities are typically two to four orders of magnitude lower than the core peak.
The core population is also modulated by the ULF wave and demonstrates anisotropy.
Right at the shock front near $x = -65\,d_{i,up}$ (Figure \ref{fig2}m-n), the ion velocity distributions illustrate a transition from the upstream to the downstream populations with reflection caused by the sharp changes in magnetic field.
Weak electrostatic perturbations are also evident in the quasi-parallel case, both upstream and downstream, as indicated by correlated electron pressure and density variations (Figure \ref{fig2}a and 2c).
Under the simulated shock conditions, the electrons are magnetized and primarily subject to adiabatic heating, a characteristic also observed in MMS case studies (e.g., \citet{khotyaintsev2024ion}).
Electrons in the downstream and at the shock display isotropic distributions (Figure \ref{fig2}k-l), while electrons in the upstream exhibit slight anisotropy due to the presence of an additional reflected electron population (Figure \ref{fig2}o-p, s-t).

Within the upstream region ($x > -60\,d_{i,\text{up}}$), two distinct wave modes are clearly identified (Figure \ref{fig2}a, \ref{fig2}d, and \ref{fig2}e).
The first mode consists of large-scale, linearly polarized waves with a wavelength of $\lambda \sim 40\,d_{i,\text{up}} \sim 5,300\,\text{km}$, a period of $T_\text{w} \sim 2\,(2\pi\omega_{ci,\text{up}}^{-1}) \sim 26\,\text{s}$, and a wave amplitude of $\delta B / B_0 \simeq 1$.
These waves extend far upstream.
The second mode comprises small-scale, left-hand polarized (in the lab frame) waves with a wavelength of $\lambda \lesssim 3.8\,d_{i,\text{up}} \sim 500\,\text{km}$, a period of $T_\text{w} \sim 1/4\,(2\pi\omega_{ci,\text{up}}^{-1}) \sim 3\,\text{s}$, and a wave amplitude of $\delta B / B_0 \lesssim 0.5$.
These waves are localized on the leading edges of steepened structures.
While both wave types exhibit magnetic field perturbations, only the large-scale mode induces significant in-phase density perturbations (Figure \ref{fig2}a and \ref{fig2}d).
Analysis of their phase velocities reveals that both wave modes propagate slightly slower than the inflow velocity in the lab frame.
Transforming to the plasma rest frame (i.e., the frame moving with the inflow velocity), the phase velocities reverse direction, now pointing upstream in the +x direction.
Importantly, the left-hand polarization of the small-scale waves in the lab frame becomes right-hand polarized in the plasma rest frame.
Based on these polarization and propagation characteristics, we identify the large-scale mode as the fast magnetosonic mode and the small-scale mode as the whistler mode, which in fact lies in the same branch in the cold plasma dispersion relation \citep{gary1993theory}.
These magnetosonic-whistler waves are triggered by instabilities corresponding to the ion-ion beam interactions \citep{gary1984electromagnetic}.

%YUXI: \textcolor{red}{(Just curious, is the wavelength sensitive to the mi/me ratio? We do not have to discuss this in the paper. I do don't know why, but it seems my 1D simulations show the shocklet structures better than Figure 2.)}

% As the fast magnetosonic waves are convected back towards the shock by the super-Alfvénic upstream flow, they evolve into \textit{shocklets} \citep{hoppe1981upstream, wilson2016low}, i.e. steepened flanks characterized by small-scale fluctuations ($\delta B / B_0 < 2$) that function as miniature shocks. 
% \textit{Short, Large Amplitude Magnetic Structures} (SLAMS, or magnetic pulsations, \citet{schwartz1992observations}) share similarities with shocklets but are defined by larger magnetic field fluctuations ($\delta B / B_0 > 2$) and potential soliton-like behavior.
% Consequently, our nonlinearly steepened ULF wave signatures align more closely with observed shocklets, where nonlinear evolution is driven by interactions with energetic particle pressure gradients, and the ensemble of these structures forms the shock transition (e.g., \citep{dubouloz1995two}).
At this snapshot $t = 19\,\omega_{ci,\text{up}}^{-1}$, a shocklet has just arrived at the main shock front.
The normal velocity between $x=-65\,d_{i,\text{up}}$ and $-55\,d_{i,\text{up}}$ exhibits a step-like profile between upstream and downstream values, accompanied by a nearby density peak.
Notably, whistler-like velocity and magnetic field fluctuations are identified in the foreshock region near the main shock front and trailing edges of density enhancements.
The whistler critical Mach number depends on $m_i / m_e$ that affects the precursor existence condition.
Larger mass ratios makes it easier to satisfy the criterion and trigger whistlers.
Section \ref{sssec:mass-ratio} discusses on this point further.

% It is crucial to acknowledge certain limitations inherent in the 1D planar shock setup.
% Firstly, with only one spatial degree of freedom, wave propagation is constrained to the x-direction.
% We established our background upstream magnetic field at a shock-normal angle $\theta_{Bn} = 30^\circ$ relative to the -x direction.
% Consequently, this oblique wave angle remains fixed and may not align with the maximum wave growth direction predicted by linear theory.
% However, \citet{lucek2002parallel} reported from in-situ satellite measurements that the average $\theta_{Bn}$ is approximately $20^\circ$, a value inconsistent with the prediction for maximum parallel propagating linear wave growth.
% This discrepancy highlights the complexity of wave behavior in real-world environments compared to idealized simulations.
% Secondly, the periodic condition applied in the y and z direction limits the growth of structures that are intrinsically high-dimensional, such as downstream plasma jets and return flows.
% Further discussion with 2D results is presented in Section \ref{sssec:2dqpar}.

\subsubsection{Impact of Mass Ratio}\label{sssec:mass-ratio}

\begin{figure}
\noindent\includegraphics[width=\textwidth]{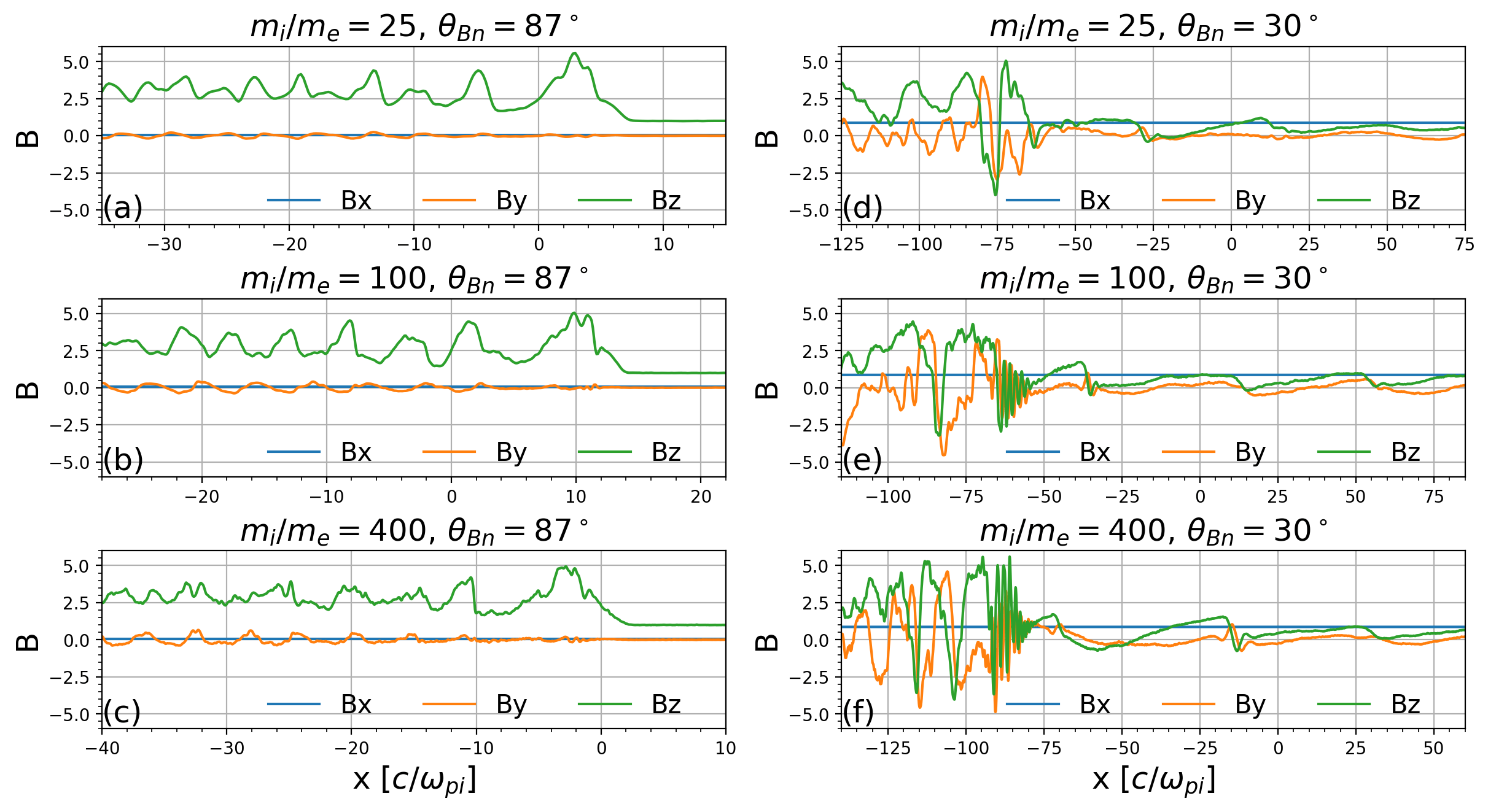}
\caption{Influence of the ion-to-electron mass ratio $m_i/m_e$ on the normalized magnetic field profile across 1D collisionless shocks at $t=20\,\omega_{ci,\text{up}}^{-1}$. The grid resolution $\Delta x = 1\,d_e$. Panels (a-c) show quasi-perpendicular shocks ($\theta_{Bn}=87^\circ$), while panels (d-f) show quasi-parallel shocks ($\theta_{Bn}=30^\circ$). The x-axes are shifted to align the shock fronts for easier comparison. The spatial extent shown is $50\,d_{i,\text{up}}$ for quasi-perpendicular shocks and $200\,d_{i,\text{up}}$ for quasi-parallel shocks.}
\label{fig-B-compare}
\end{figure}

To assess the sensitivity of 1D shock structures to the ion-to-electron mass ratio $m_i/m_e$, we conduct simulations with a fixed grid resolution relative to the electron skin depth $d_e$, specifically $\Delta x = d_e$.
We keep the real ion mass while setting different electron masses in the simulations.
Figure \ref{fig-B-compare} compares the magnetic field profiles for quasi-perpendicular (a-c) and quasi-parallel (d-f) shocks with $m_i/m_e = 25,\,100,\,400$.
Because the grid resolution is fixed with respect to $d_e$, the effective resolution in terms of the ion inertial length $d_i$ varies with the mass ratio as $\Delta x = d_e = d_i\sqrt{m_e/m_i}$, resulting in a corresponding change in the number of cells across the shock transition.
Despite this difference in effective $d_i$ resolution, the overall magnetic field profiles remain remarkably similar across the different mass ratios, in both field perturbation magnitudes and periods.

However, a key distinction arises in the presence of whistler precursors in the quasi-parallel $\theta_{Bn}=30^\circ$ cases.
In the low mass ratio $m_i/m_e = 25$ run (Figure \ref{fig-B-compare}d), the nonlinear steepening does not associate with obvious whistler precursors.
We observe clear whistler waves upstream of the shock front with wavelength $\lambda_w \sim 2d_{i,\text{up}}$ in larger mass ratio cases (Figure \ref{fig-B-compare}e and f), where the whistler critical Mach number $\text{M}_\text{w}\gtrsim \text{M}_\text{A}$, consistent with the formation criterion (Equation \ref{eq-whistler-mach}).
It is also noted that the local Mach number is modified in the presence of waves, which potentially makes it easier to generate precursors in the quasi-parallel runs.
Similar whistler precursor occurrences are reported in full PIC simulations, e.g. \citet{tsubouchi2004slams,  nakanotani2022collisional}.
These whistler waves consistently appear on the leading edges in the first few ULF wavelengths ahead of the shock, which may be linked directly to the main shock front or the upstream steepened shocklets \citep{wilson2016low}.

%Proton-to-electron mass ratio may also affect the cross-shock energy partition as well as wave dispersion relations.

\subsubsection{Effect of Grid Resolution}

\begin{figure}
\noindent\includegraphics[width=\textwidth]{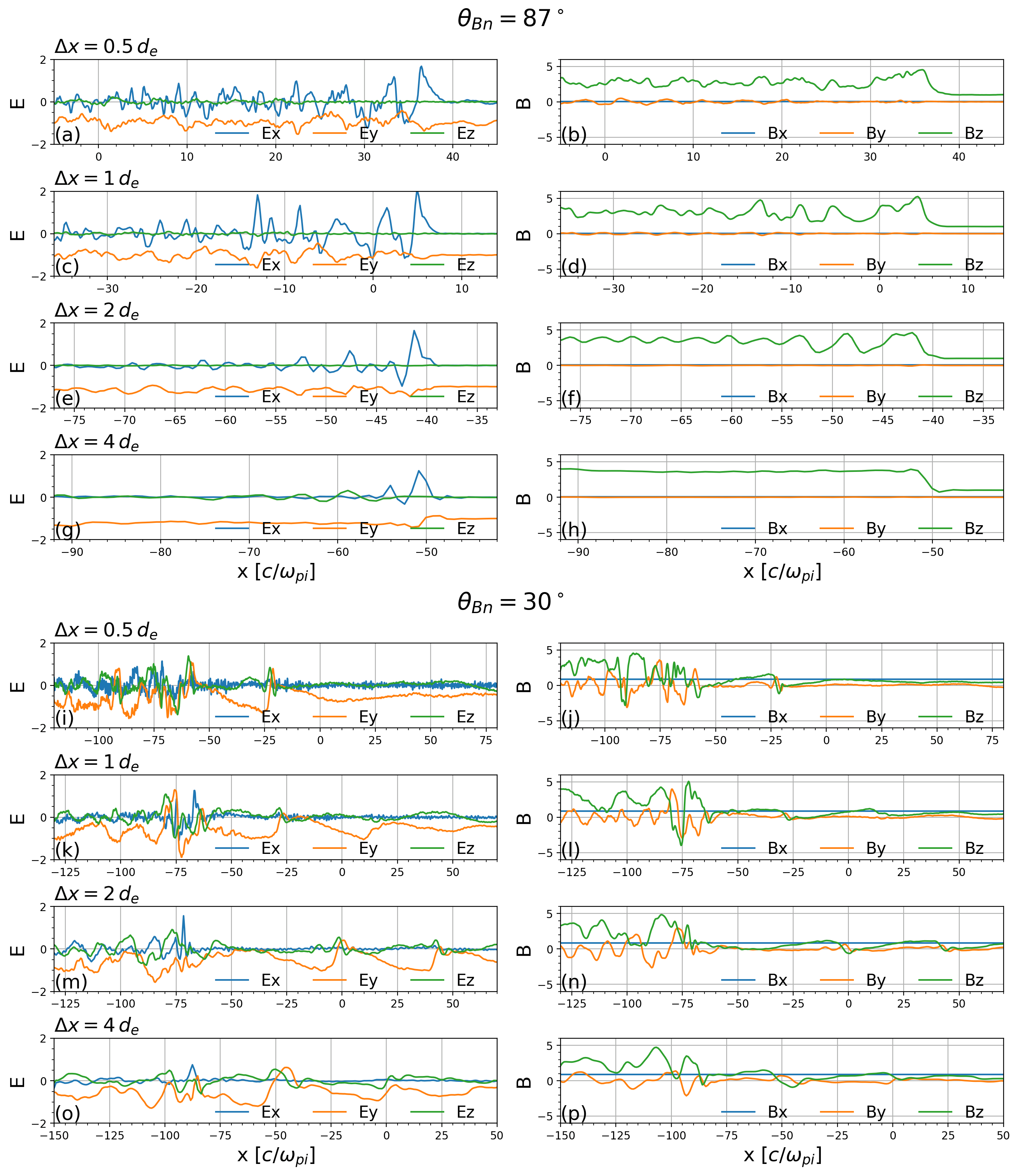}
\caption{Influence of the grid resolution $\Delta x$ on the normalized electromagnetic field profile across 1D collisionless shocks of $m_i / m_e =25$. Panels (a-h) show quasi-perpendicular shocks ($\theta_{Bn}=87^\circ$) at $t=20\,\omega_{ci,\text{up}}^{-1}$, while panels (i-p) show quasi-parallel shocks ($\theta_{Bn}=30^\circ$) at $t=20\,\omega_{ci,\text{up}}^{-1}$. The x-axes are shifted to align the shock fronts for easier comparison. The spatial extent shown is $50\,d_{i,\text{up}}$ for quasi-perpendicular shocks and $200\,d_{i,\text{up}}$ for quasi-parallel shocks.}
\label{fig-res-compare}
\end{figure}

To investigate the influence of grid resolution on the simulation results, we conducted a series of 1D simulations with a fixed mass ratio $m_i/m_e = 25$ and varying grid resolutions $\Delta x$.
Figure \ref{fig-res-compare} compares the electromagnetic field profiles for fully developed quasi-perpendicular (a-d) and quasi-parallel (e-h) shocks, respectively, with $\Delta x = 0.5,\,1,\,2,\,4\,d_{e,\text{up}}$.
The coarsest resolution $\Delta x = 4\,d_{e,\text{up}}\simeq 1\,d_{i,\text{up}}$ is close to the ion inertial length.
In the quasi-perpendicular shock simulations, coarser resolutions lead to smoother EM field profiles, while the solutions stay meaningful at sub-ion scales.
Similarly, for quasi-parallel shocks, coarser resolutions result in a weakening of the downstream and upstream field fluctuations, but large-scale structures persist.
These observations demonstrate the asymptotic behavior in \texttt{FLEKS} shock simulations across the sub-ion scales.

\subsection{Local 2D Simulations}\label{ssec:2d}

To further validate the model, we extend the investigation to two dimensions (2D).
These 2D simulations employ the same initial plasma parameters and proton-to-electron mass ratio of $m_i / m_e = 25$ as the 1D cases, with the shock front initially normal to the x-direction and the magnetic field lying in the x-y plane.
The simulation domain spans $-70\,d_{i,\text{up}}$ to $70\,d_{i,\text{up}}$ in the y direction.
The x-extent is $200\,d_{i,\text{up}} \simeq 26,400\,\text{km}$ for the quasi-perpendicular shock case and $400\,d_{i,\text{up}} \simeq 52,800\,\text{km}$ for the quasi-parallel shock case, consistent with the 1D runs.

\subsubsection{2D Quasi-Perpendicular Shock}\label{sssec:2dqperp}

\begin{figure}
\noindent\includegraphics[width=\textwidth]{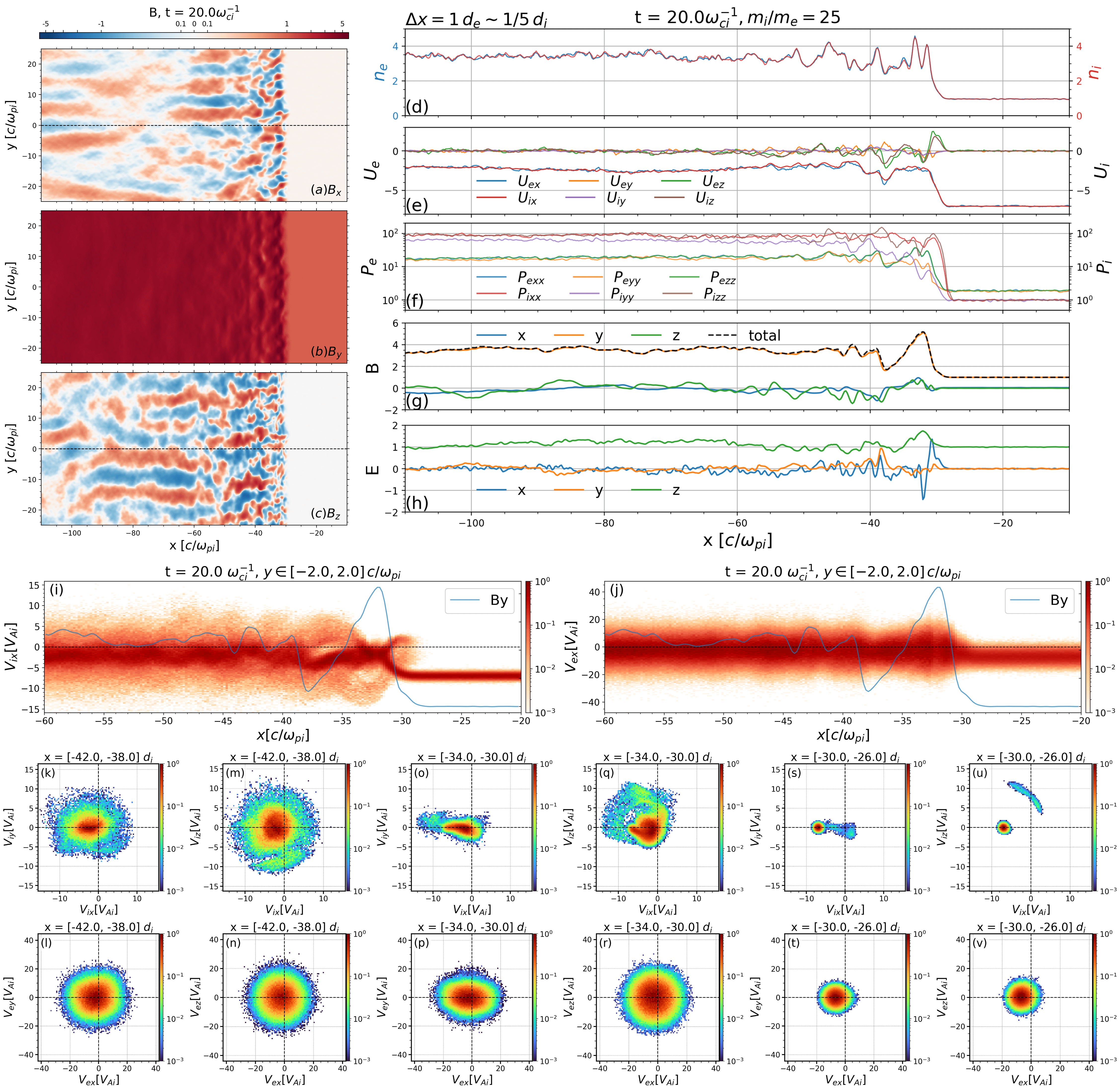}
\caption{2D $\theta_{Bn} = 87^\circ$ quasi-perpendicular shock run with $m_i / m_e = 25$ at $t=20\,\omega_{ci,\text{up}}^{-1}$. The plasma moments and EM fields are normalized by the upstream quantities. Phase space distributions are normalized by the maximum phase space density in the displayed regions. (a-c) Zoomed-in view of magnetic field components near the shock front. (d-h) Plasma quantities and electromagnetic fields along the y=0 dashed line indicated in panel a. (i,j) Ion and electron $x-v_{x}$ phase space plots, with the $B_y$ profile shown in blue for reference. (k-n)  Ion and electron velocity space distributions in the downstream region. (o-r) Ion and electron velocity space distributions at the shock front. (s-v) Ion and electron velocity space distributions ahead of the shock ramp. The y-extent for taking the velocity distributions goes from $-2\,d_{i,\text{up}}$ to $2\,d_{i,\text{up}}$.}
\label{fig3}
\end{figure}

Figure \ref{fig3} presents a detailed view of plasma moments, electromagnetic fields, and electron and ion phase space distributions near the 2D quasi-perpendicular $\theta_{Bn} = 87^\circ$ shock at $t=20\,\omega_{ci,,\text{up}}^{-1}$, with the shock front located around $x=-30\,d_{i,\text{up}}$.
In contrast to the 1D supercritical case (Figure \ref{fig1}), the 2D simulation reveals a more dynamic downstream region characterized by distinct wave activity, while the upstream region remains steady and quiescent.
This increased dynamism is partly due to more efficient gyrophase mixing in the 2D geometry with in-plane background magnetic field, as evidenced by the faster decay of downstream oscillations compared to the 1D case.

Remnants of the 1D downstream periodic structures are still discernible between $x = -50\,d_{i,\text{up}}$ and $-30\,d_{i,\text{up}}$ (Figure \ref{fig3}d, \ref{fig3}f, \ref{fig3}g), but their extent in the shock normal direction is limited to less than $20\,d_{i,\text{up}}$.
The normalized magnetic fields in Figure \ref{fig3}a-c are presented on a symmetric logarithmic scale centered at 0.
Immediately downstream of the shock front at $x=-30\,d_{i,\text{up}}$, strong fluctuations in the transverse $B_x$ and $B_z$ components are shown in the color contours, with amplitudes $\lesssim 2\,B_{\text{up}}$ and a wavelength of approximately $4\,d_{i,\text{up}}$. 
The comparable magnitudes of $B_x$ and $B_z$ oscillations suggest near-circular polarization, identifying these fluctuations as the parallel-propagating Alfvén ion-cyclotron (AIC) waves, also known as the electromagnetic ion cyclotron (EMIC) waves.
These waves are generated by ion anisotropy shown in Figure \ref{fig3}o-p, and also contributes to the observed surface rippling along the y-direction.

Further downstream, the field fluctuations mainly occur in the $B_z$ component, indicating a transition from parallel to oblique wave propagation, similar to the hybrid simulation results presented in \citet{lee2017generation}, Section 4.
The fluctuations along the x-direction corresponds to the existence of mirror modes, with nearly zero phase velocity in the lab frame (i.e. moving with the bulk plasma).
Mirror modes are driven by downstream anisotropic ion distributions, and may also be identified via the anti-correlations between plasma density and total magnetic field (Figure \ref{fig3}d and \ref{fig3}g) at $x\simeq -110, -70\,d_{i,\text{up}}$).
%At $x\simeq -110, -70\,d_{i,\text{up}}$), anti-correlations between plasma density and total magnetic field emerge (Figure \ref{fig3}d and \ref{fig3}g), a signature of mirror modes driven by the anisotropic ion distributions.
These are not classified as MHD slow modes because of the lack of significant velocity perturbations \citep{song1994identification}.
We should note that mirror modes can arise outside the $y=0$ line cut, a possibility that is not captured by examining only this specific slice.
The mirror modes can also partly be reflected from the compressive magnetic field perturbation $\delta B_\parallel \simeq \delta B_y$ in Figure \ref{fig3}b.

%Gabor: \textcolor{red}{(The reviewer will ask you to look at other slices. Better do now)}

A survey of the downstream region reveals that the ion and electron distributions remain relatively consistent across different locations, represented by those shown in Figure \ref{fig3}k-n.
At the shock front (Figure \ref{fig3}o-p), the ion velocity distribution transitions from a single cold Maxwellian upstream to a mixture of two Maxwellian distributions and a curved diffusion path in the $V_{ix}-V_{iz}$ plane.
Near the foot region (Figure \ref{fig3}s-t), minimal ion reflection is observed, with reflected ions undergoing acceleration along the tangential z-direction.
In contrast, electrons are preferentially heated in the perpendicular plane (Figure \ref{fig3}m and \ref{fig3}q) and exhibit gyrotropic trajectories (Figure \ref{fig3}n and \ref{fig3}r).
They become slightly anisotropic with $P_{i\perp}/ P_{i\parallel} \lesssim 2$, starting immediately at the shock front.
However, these magnetized, thermalized electrons do not significantly interact with the ion-scale EMIC and mirror mode waves.

In contrast to the 1D simulation, the 2D setup, with the background magnetic field oriented primarily along the y-direction in-plane, allows for proper parallel heating and the release of free energy associated with downstream temperature anisotropies (Figure \ref{fig3}k-n).
To assess the potential for instabilities driven by these anisotropies, we consider the linear instability criteria for mirror modes and EMIC waves.

The linear instability criterion for mirror modes is given by \citet{southwood1993mirror}:
\begin{linenomath*}
\begin{equation} \label{eq-mirror}
\left( \frac{P_{i\perp}}{P_{i\parallel}} - 1 \right) \beta_{i\perp} > 1
\end{equation}
\end{linenomath*}
where $P_{i\perp}$ and $P_{i\parallel}$ are the perpendicular and parallel ion thermal pressures, and $\beta_{i\perp}=P_{i\perp}/P_B$.
In the downstream region, the ion anisotropy saturates at $P_{i\perp}/ P_{i\parallel} \simeq (P_{i,xx} + P_{i,zz}) / 2 P_{i,yy} \simeq 1.5$, and $\beta_{i\perp} = P_{i,\perp} / P_B \simeq \mu_0 (P_{i,xx} + P_{i,zz}) / B_\text{down}^2 \sim 2.8$. These values clearly satisfy the mirror instability criterion (\ref{eq-mirror}).

Similarly, the linear instability criterion for EMIC waves is given by \citet{gary1993theory}:
\begin{linenomath*}
\begin{equation} \label{eq-emic}
\left( \frac{P_{i\perp}}{P_{i\parallel}} - 1 \right) \frac{S}{\beta_{i\parallel}^\alpha} > 1
\end{equation}
\end{linenomath*}
where $\beta_{i\parallel}=P_{i\parallel}/P_B$, $S$ and $\alpha$ are fitting parameters dependent on the wave growth rate $\gamma$ and plasma conditions. 
For simplicity, we adopt the fitting parameters from \citet{blum2012comparison} for $\gamma/\omega_{ci,\text{up}} = 0.001$, yielding $S=0.429$ and $\alpha=0.409$. With $\beta_\parallel \simeq 2\mu_0 P_{i,yy} /B_\text{down}^2 \sim 1.9$, the EMIC instability criterion (\ref{eq-emic}) is also marginally satisfied.
This linear instability analysis confirms that the conditions downstream of the 2D quasi-perpendicular shock are conducive to the generation of both mirror modes and EMIC waves.
% The 2D simulation, incorporating realistic solar wind conditions upstream of Earth \citep{salem2023precision}, provides a more accurate representation of the plasma environment at Earth's bow shock compared to the 1D case.
% This highlights the importance of at least 2D modeling for capturing the complex dynamics of quasi-perpendicular shocks, including the interplay of anisotropy-driven instabilities and wave generation.

\subsubsection{2D Quasi-Parallel Shock}\label{sssec:2dqpar}

\begin{figure}
\noindent\includegraphics[width=\textwidth]{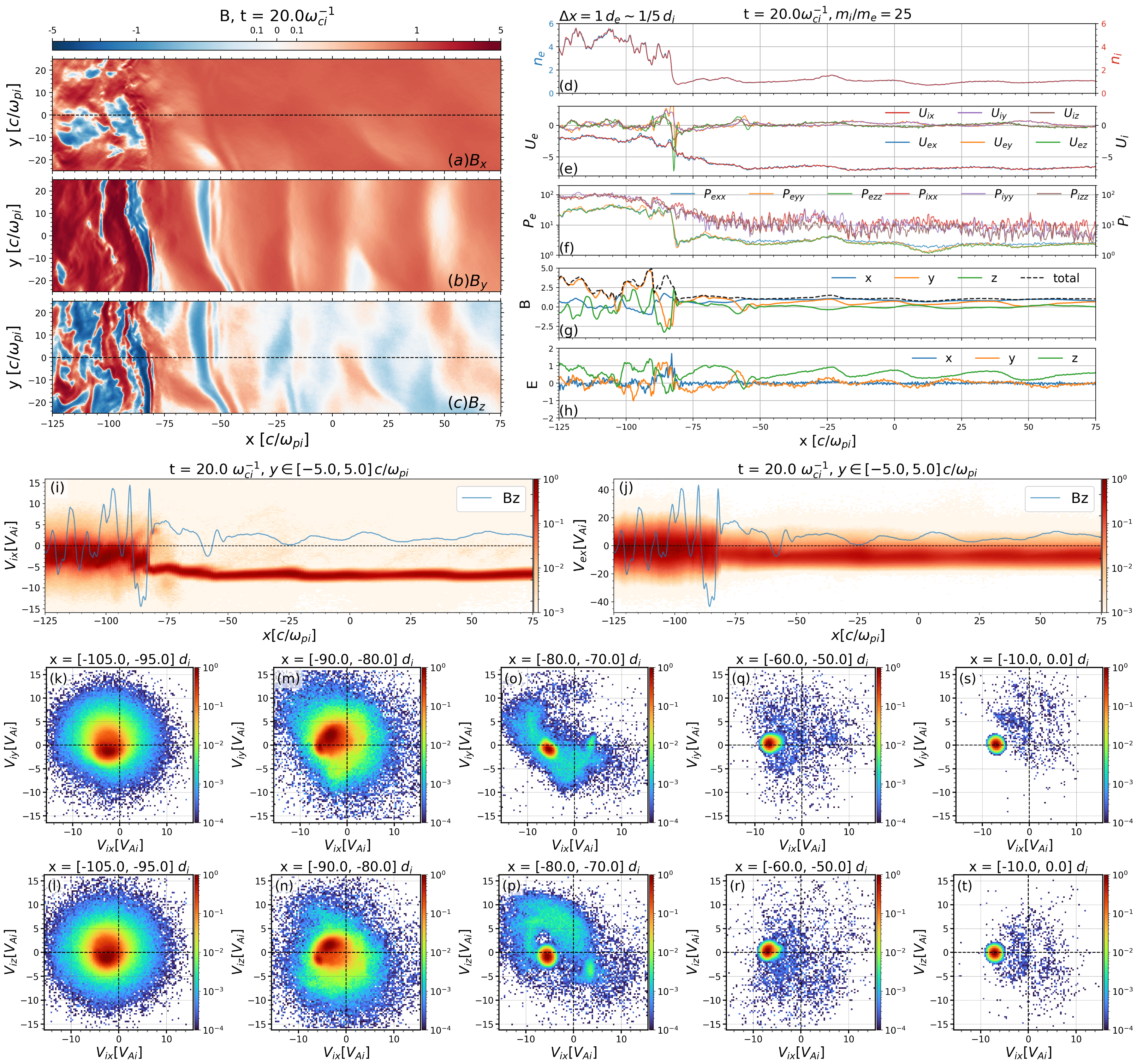}
\caption{2D $\theta_{Bn}=30^\circ$ quasi-parallel shock run with $m_i / m_e = 25$ at $t=20\,\omega_{ci,\text{up}}^{-1}$. The plasma moments and EM fields are normalized by the upstream quantities. Phase space distributions are normalized by the maximum phase space density in the displayed regions. (a-c) Zoomed-in view of magnetic field components near the shock front. (d-h) Plasma quantities and electromagnetic fields along the $y=0$ dashed line indicated in panel a. (i-j) Ion and electron $x-V_x$ phase space density. (k-t) Ion velocity space distributions at five locations across the shock front, each spanning $10\,d_{i,\text{up}}$ in the x-direction and sampled over $-5\,d_{i,\text{up}} \leq y \leq 5\,d_{i,\text{up}}$.
}
\label{fig4}
\end{figure}

In the 2D quasi-parallel shock simulation, the upstream magnetic field is oriented at a $30^\circ$ angle to the +x direction, tilted towards the +y direction in-plane.
Initially the out-of-plane $B_z$ is zero.
Figure \ref{fig4} presents a detailed view of the 2D quasi-parallel shock results, following a similar format to Figure \ref{fig3}.
The main shock front, located near $x=-80\,d_{i,\text{up}}$ at $t = 20\,\omega_{ci,\text{up}}^{-1}$, exhibits surface waves along the tangential direction, distorting the initially planar shock surface. 
Unlike the 1D simulations, variations in $B_x$ are now present (Figure \ref{fig4}a).
The upstream wave fronts are also tilted due to shock rippling (see Figure~\ref{fig7}).
Similar to the 1D case, the downstream region exhibits stronger, more irregular magnetic field fluctuations than the upstream region (Figure \ref{fig4}b-c).
The dominant wave mode observed upstream is the large-scale fast magnetosonic wave, with wavelength $\lambda \sim 50\,d_{i,\text{up}} \sim 6,600\,\text{km}$, period $T_\text{w} \sim 1.5\,(2\pi\omega_{ci,\text{up}}^{-1}) \sim 20\,\text{s}$ and $\delta B / B_0 \simeq 1$, where $\delta B$ is the fluctuating magnetic field and $B_0$ is the background average magnetic field.
Whistler-like velocity and magnetic field perturbations take place at the main shock front and the leading edges of steepened structures.

%Two distinct wave modes are observed upstream: large-scale fast magnetosonic waves (wavelength $\lambda \sim 50\,d_{i,\text{up}} \sim 6,600\,\text{km}$, period $T_\text{w} \sim 1.5\,(2\pi\omega_{ci,\text{up}}^{-1}) \sim 20\,\text{s}$, and $\delta B / B_0 \simeq 1$, where $\delta B$ is the fluctuating magnetic field and $B_0$ is the background average magnetic field) and small-scale whistler waves (wavelength $\lambda \sim 3\,d_{i,\text{up}} \sim 400\,\text{km}$, period $T_\text{w} \sim 1/5\,(2\pi\omega_{ci,\text{up}}^{-1}) \sim 2.6\,\text{s}$, $\delta B / B_0 \lesssim 0.5$) located on the leading edges of steepened structures.

Figure \ref{fig4}d-h shows the plasma moments and electromagnetic field profiles along the $y=0$ cut indicated by the black dashed line in Figure \ref{fig4}a.
These profiles closely resemble the 1D profiles in Figure \ref{fig1}a-e, with the exception that the y and z components of the fields are switched due to the initial magnetic field lying in the x-y plane in this 2D simulation.
However, note that the line profiles at different y cuts can be different with variations along the y-direction.

The ion $x-V_{ix}$ phase space plot, Figure \ref{fig4}i, reveals a reflected ion beam extending approximately $5\,d_{i,\text{up}}$ upstream, with a beam density of about 3\% of the inflow ion density, surrounded by a diffuse ion population.
The free energy from magnetic and pressure gradients in the suprathermal ions can lead to further wave amplification and steepening (e.g. \citet{scholer2003short}).
Meanwhile, electrons shown by the $x-V_{ex}$ phase space plot Figure \ref{fig4}j remain magnetized and are primarily heated adiabatically.

Figure \ref{fig4}k-t presents the ion velocity distribution projections onto the $V_x-V_y$ and $V_x-V_z$ planes at five different locations, ordered from downstream to upstream.
In the downstream region near $x=-100\,d_{i,\text{up}}$ (Figure \ref{fig4}k-l), ions exhibit approximately a thermal Maxwellian distribution with a drift in the -y direction.
Closer to the shock front near $x=-85\,d_{i,\text{up}}$ (Figure \ref{fig4}m-n), a transition from the upstream to the downstream ion population takes place.
Right in front of the shock around $x=-75\,d_{i,\text{up}}$ (Figure \ref{fig4}o-p), a clear reflected beam is observed, centered at $(V_{ix}, V_{iy}, V_{iz}) = (2, 1, -3)\,V_{Ai,\text{up}}$ and with a beam velocity of approximately $10\,V_{Ai,\text{up}}$.
A diffuse ion population is also present across a wide range of velocities.
Further upstream in the foreshock region, the reflected diffuse ion population aligns with the upstream magnetic field, coexisting with the core Maxwellian inflow (Figure \ref{fig4}q-r).
The diffuse ion population becomes sparser deep in the foreshock (Figure \ref{fig4}s-t), but can be found all the way to the inflow domain boundary.

\begin{figure}
\noindent\includegraphics[width=\textwidth]{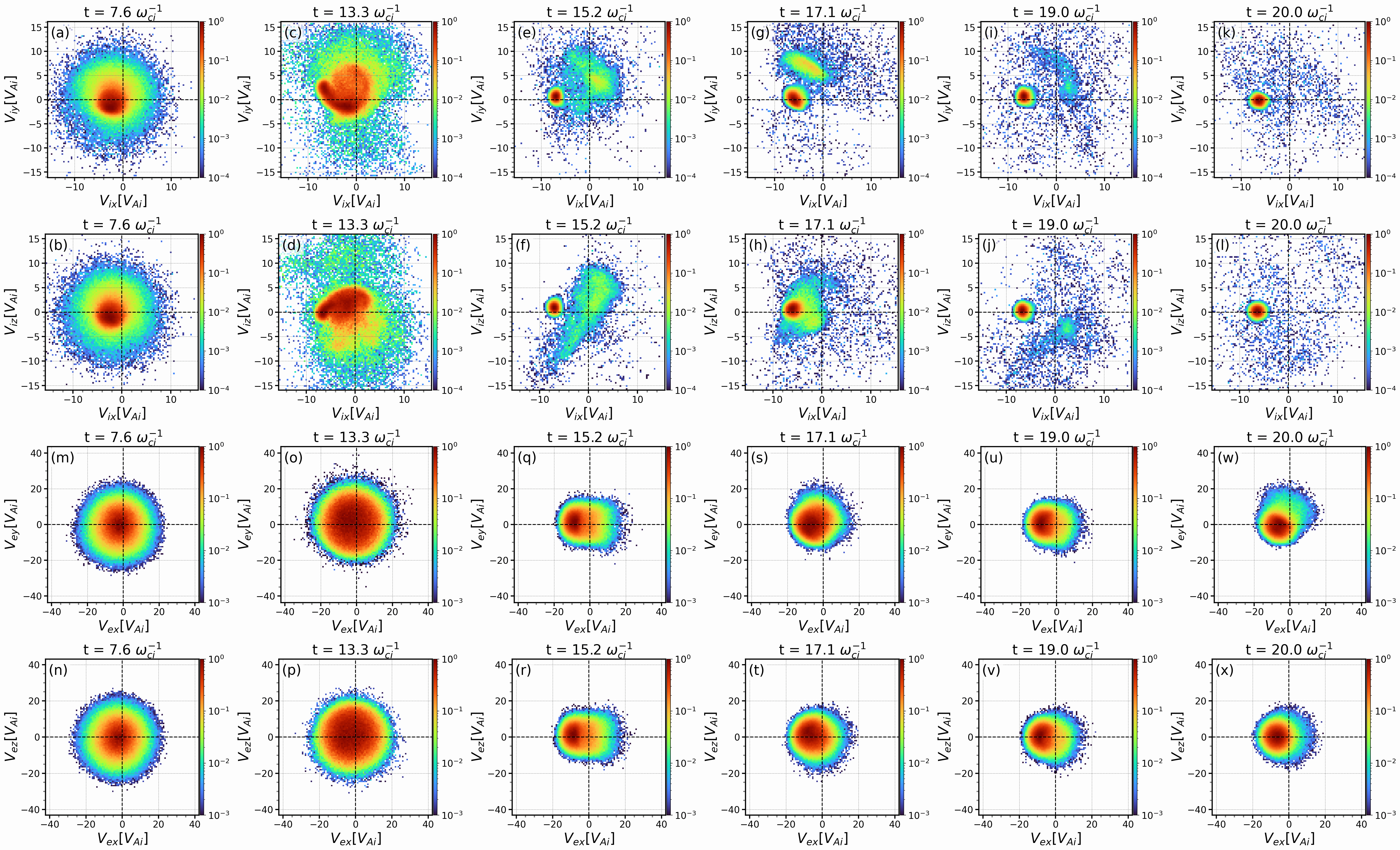}
\caption{Time evolution of the ion (a-l) and electron (m-x) $V_x-V_y$ and $V_x-V_z$ velocity space densities at $x \in [-65, -60]\,d_{i,\text{up}}$, $y \in [-2.5, 2.5]\,d_{i,\text{up}}$ of the 2D $\theta = 30^\circ$ quasi-parallel shock run from $t = 7.6$ to $20\,\omega_{ci,\text{up}}^{-1}$. Each column is taken from the same snapshot. Phase space distributions are normalized by the maximum phase space density at each snapshot.}
\label{fig5}
\end{figure}

The simulated quasi-parallel shock is not steady nor in an equilibrium state, with the shock front moving gradually along the x-direction.
We next check the particle behaviors at a fixed location to account for the spatiotemporal evolution.
Figure \ref{fig5} displays the time evolution of ion and electron distributions within a $5\,d_{i,\text{up}} \times 5\,d_{i,\text{up}}$ box centered at $x=-62.5\,d_{i.\text{up}}$.
This location, initially downstream of the shock, transitions to the upstream region as the shock front propagates in the -x direction.
Each column represents a snapshot in time, while each row shows the velocity space density projections onto the $V_{x}-V_{y}$ and $V_x-V_z$ planes for ions and electrons, respectively.
At $t = 0\,\omega_{ci,\text{up}}^{-1}$, the plasma is initialized with single Maxwellian distributions based on MHD solutions.
By $t = 7.6\,\omega_{ci,\text{up}}^{-1}$, the transition to a kinetic state is evident, with the emergence of a non-Maxwellian but nearly isotropic ion distribution.
At $t = 13.3\,\omega_{ci,\text{up}}^{-1}$, the shock front passes by, shown by the central drift path in Figure\ref{fig5}c-d.
A distinct ion beam propagating upstream forms by $t = 15.2\,\omega_{ci,\text{up}}^{-1}$, centered at $V_{ibx}\simeq 3\,V_{Ai,\text{up}}$.
This field-aligned beam, with a relative velocity of $V_0 = V_{0b} - V_{0m} \simeq 10\,V_{Ai,\text{up}}$ with respect to the upstream Maxwellian population, is a characteristic feature of the foreshock region. 
However, this beam structure is transient and also affected by the local magnetic field orientation (Figure \ref{fig5}g-h).
By $t = 19\,\omega_{ci,\text{up}}^{-1}$, it has dissipated, leaving a more diffuse and sparse upstream-moving population afterwards (Figure \ref{fig5}k-l).
%This observation is consistent with Figure \ref{fig4}i, where the temporary reflected ion beam spans approximately $5\,d_{i,\text{up}}$.
As the shock front continues to propagate in the -x direction, this fixed spatial location near $x=-62.5\,d_{i,\text{up}}$ samples the diffuse ion population that was previously further upstream (Figure \ref{fig5}g-h).
The phase space densities of this diffuse population are typically below 1\% of the peak density in the upstream core population.
By the final snapshot at $t=20\,\omega_{ci,\text{up}}^{-1}$ (Figure \ref{fig5}i-j), the diffuse ion population becomes even more sparse.
This time evolution complements the spatial comparisons of ion distribution in Figure \ref{fig4}k-t, providing a comprehensive picture of the spatiotemporal dynamics across the quasi-parallel shock.

The lower half of Figure \ref{fig5} shows the corresponding time evolution of the electron velocity distribution.
As the shock front approaches (Figure \ref{fig5}m-p), the electron temperature increases while the distribution remains Maxwellian.
At later times $t>15\,\omega_{ci,\text{up}}^{-1}$, when the observation location moves into the foreshock region, the electron distribution becomes non-Maxwellian with diffuse reflected species.
This non-equilibrium distribution indicates a source of free energy that can potentially drive the small-scale waves observed in conjunction with the large-scale ULF magnetosonic waves.
These waves, if exist, are likely triggered by sub-ion-scale instabilities, distinct from the ion-ion instability responsible for the magnetosonic waves \citep{gary1993theory}.
However, in this planar shock setup with low mass ratio $m_i / m_e = 25$ and $\Delta x = 1\,d_e$, the sub-ion scale waves are not obvious.

\begin{figure}
\noindent\includegraphics[width=\textwidth]{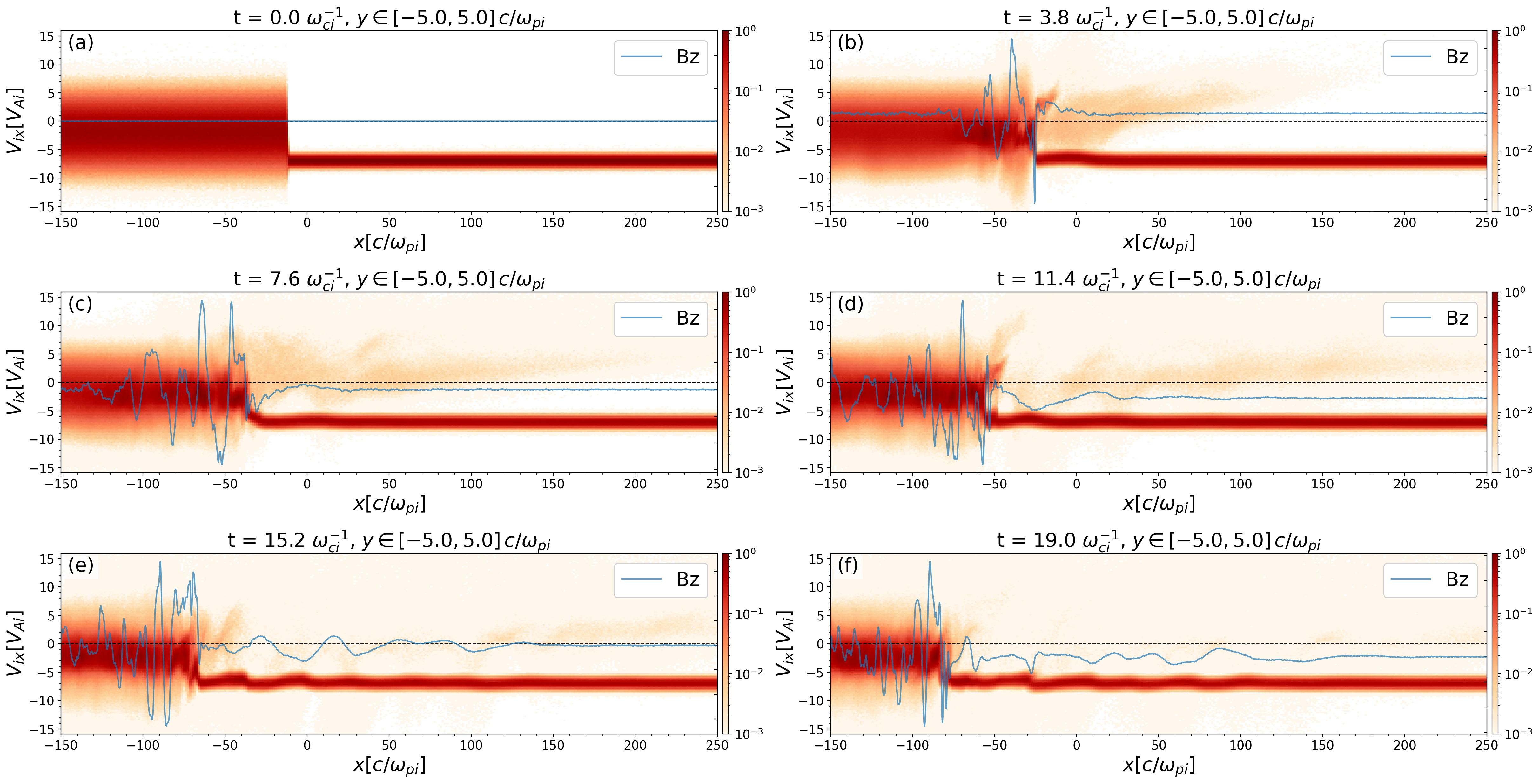}
\caption{Time evolution of the ion $x-V_{ix}$ phase space densities for $x \in [-150, 250]\,d_{i,\text{up}}$, $y \in [-5, 5]\,d_{i,\text{up}}$ of the 2D $\theta = 30^\circ$ quasi-parallel shock run at six snapshots from $t = 0$ to $19\,\omega_{ci,\text{up}}^{-1}$. The magnetic field $B_z$ profiles along the $y=0$ cut are shown in blue lines as references. Phase space distributions are normalized by the maximum phase space density at each snapshot.}
\label{fig6}
\end{figure}

To track the time evolution of reflected ions, we again look at the $x-V_{ix}$ ion phase space distributions presented in Figure \ref{fig6} from $t = 0$ to $19\,d_{i,\text{up}}$.
The whole simulation domain in the x-direction from $-150\,d_{i,\text{up}}$ to $250\,d_{i,\text{up}}$ is shown to facilitate a detailed analysis of the foreshock region.
The initial MHD state (Figure \ref{fig6}a) contains no reflected ions before encountering the shock.
As the kinetic system evolves, reflected ions appear early in the simulation (Figure \ref{fig6}b) and rapidly populate the upstream region (Figure \ref{fig6}c).
However, the distinct beam-like structures extend only up to $\lesssim 10\,d_{i,\text{up}}$ ahead of the shock and are not present in all snapshots.
Intriguingly, the growth of ULF magnetosonic waves is observed to commence after the formation of reflected ion species \citep{turc2018foreshock, turc2025foreshock}.
During the temporal delay between these snapshots, the initially coherent upstream traveling beam undergoes a transition to a diffuse population, a process that correlates with the saturation of wave growth.
In parallel, the core species of the inflow are subject to subtle modulations resulting from the ULF wave perturbations.

\begin{figure}
\noindent\includegraphics[width=0.8\textwidth]{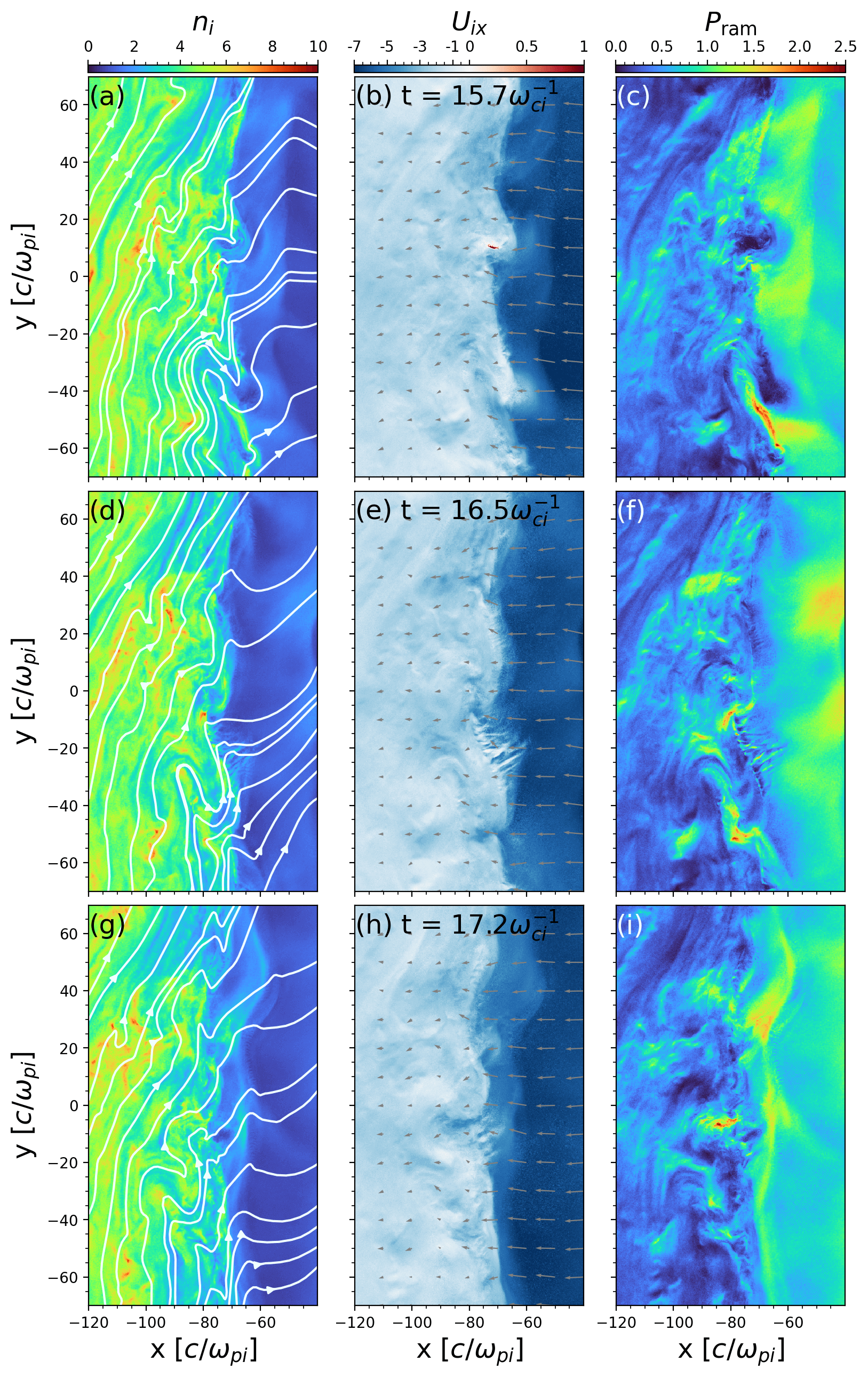}
\caption{Colored contours of ion density $n_i$ (a, d, g), velocity $U_{ix}$ (b, e, h), and ram pressure $P_\text{ram}$ (c, f, h) normalized to the upstream values close to the 2D $\theta_{Bn}=30^\circ, \mathrm{M}_\mathrm{A}=12$ quasi-parallel shock front at three consequent snapshots. The white lines in (a) are $B_x-B_y$ magnetic fields, and the gray quivers in (b) are $U_{ix}-U_{iy}$ ion velocities.}
\label{fig7}
\end{figure}

Finally we investigate the plasma moments to check the structures propagating through the quasi-parallel shock.
Figure \ref{fig7} presents a close-up view of the normalized ion density $n_i$, velocity $U_{ix}$, and ram pressure $P_\text{ram}$ near the shock front, spanning $x \in [-120, -40]\,d_{i,\text{up}}$ and $y \in [-70, 70]\,d_{i,\text{up}}$, at three consequent snapshots $t=15.7,16.5,17.2\,\omega_{ci,\text{up}}^{-1}$.
White lines with arrows and gray quivers represent in-plane magnetic field lines and ion velocities, respectively.
Surface ripples are clearly visible near $x = -70\,d_{i,up}$ across this extended tangential domain.
Several regions, both upstream and downstream, exhibit localized enhancements in ion density and velocity, leading to increased ram pressure $P_\text{ram} = m_i n_i U_i^2 / 2$.
As shown by the time series, these localized structures are associated with upstream ULF wave perturbations and the propagation of nonlinear shocklets across the shock front into the downstream region.

Of particular interest is the region with a return flow towards the upstream ($U_{ix}>0$) observed around $x=-70\,d_{i,\text{up}}$ and $y=10\,d_{i,\text{up}}$ (Figure \ref{fig7}b), highlighted using an asymmetric colormap.
This return flow reaches speeds of approximately $\simeq 1.5 V_{Ai,\text{up}}$ within a localized thin channel extending approximately $5\,d_{i,\text{up}} \simeq 650\,\text{km}$ normal to the shock surface.
The return flow likely arises from the compression of nearby high-pressure blobs that have just crossed the shock front in the y direction, as evidenced by the ram pressure gradient, similar to the satellite observation of sunward flows in the downstream magnetosheath (e.g., \citet{archer2014role}).

The localized enhancements in ram pressure are characteristic of jets.
In run 14, across a tangential width of $140\,d_{i,\text{up}} \simeq 18,000\,\text{km}$, we find an occurrence rate of approximately one jet per $2\pi/\omega_{ci,\text{up}}$, or roughly 4 jets per minute, based on the inspecting time interval from $t=9$ to $20\,\omega_{ci,\text{up}}^{-1}$.
The jets in the 2D planar shock simulations tend to dissipate rapidly in the downstream region, limiting their penetration depth to approximately $20\,d_{i,\text{up}}$.

% The localized enhancements in ram pressure are characteristic of jets \citep{plaschke2018jets}, frequently observed in the terrestrial magnetosheath downstream of quasi-parallel shocks.
% In run 14, across a tangential width of $140\,d_{i,\text{up}} \simeq 18,000\,\text{km}$, we find an occurrence rate of approximately one jet per $2\pi/\omega_{ci,\text{up}}$, or roughly 4 jets per minute, based on the inspecting time interval from $t=9$ to $20\,\omega_{ci,\text{up}}^{-1}$. 
% Despite the lack of a clear definition for jets, our simulations reveal a jet occurrence rate higher than observational estimates of a few jets per hour.
% This discrepancy might be partly attributed to the generally smaller scale of our simulated jets ($\lesssim 10\,d_{i,\text{up}}$) compared to observed terrestrial magnetosheath jets, making them more challenging for satellites to detect.
% Furthermore, in our 2D planar shock simulations, the jets tend to dissipate rapidly in the downstream region, limiting their penetration depth to approximately $20\,d_{i,\text{up}}$.
% This limited penetration is likely influenced by the downstream oblique magnetic field orientation and the inherent constraint of the two-dimensional setup, which lacks the complexity of the third spatial dimension.
% Ultimately, these fundamental differences between our simplified 2D planar shock model and the actual global 3D magnetospheric configuration offer plausible explanations for the observed model-observation mismatch.

\subsubsection{Downstream Magnetic Reconnection and Jet Dynamics in Quasi-Parallel Shocks}

\begin{figure}
\noindent\includegraphics[width=\textwidth]{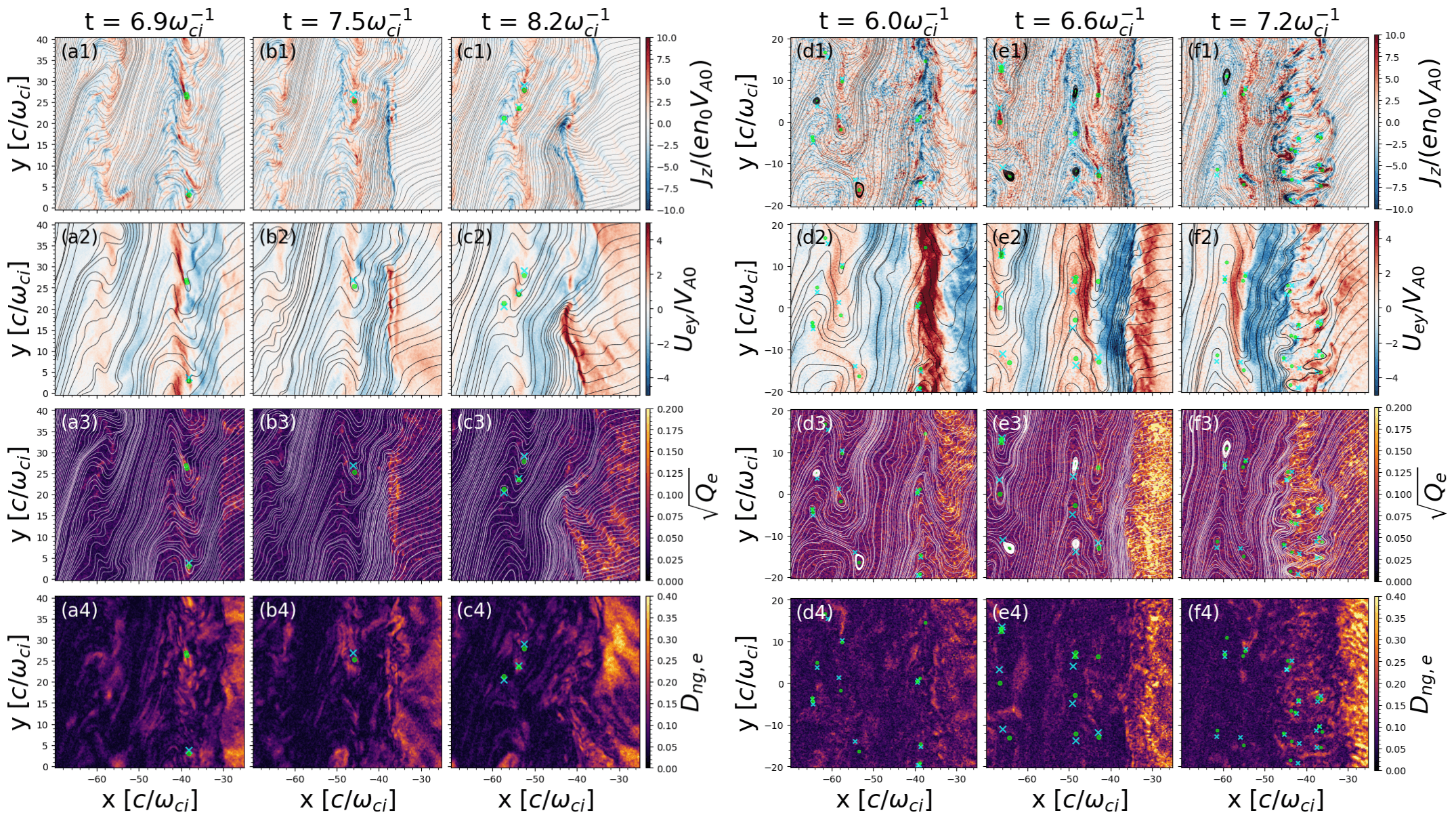}
\caption{Colored contours of out-of-plane current density $J_z$, electron velocity $U_{ey}$, and nongyrotropy measures $\sqrt{Q_e}$ and $D_{ng,e}$ in the 2D $\theta_{Bn}=30^\circ$ quasi-parallel shock run 14 ($\mathrm{M}_\mathrm{A}=7$, column a-c) and 15 ($\mathrm{M}_\mathrm{A}=12$, column d-f) at three consecutive snapshots, overplotted with in-plane magnetic field lines. Reconnection X-points and O-points are marked with cyan crosses and green circles, respectively.}
\label{fig8}
\end{figure}

The turbulent nature of the quasi-parallel shocks and the propagation of foreshock waves make the shock downstream region an ideal location for triggering magnetic reconnection \citep{wang2024origin} and forming jets \citep{ren2023two}. To resolve key kinetic processes downstream of the shock, we further perform a comparative analysis of three 2D quasi-parallel shock simulations (runs 14, 15, and 16), focusing on magnetic reconnection and jet dynamics.
We select $\sqrt{Q_e}$ and $D_\text{ng,e}$ as representatives from the list of electron non-gyrotropy measures that are enhanced at the reconnection sites \citep{zhou2020reconnection}.
Figure \ref{fig8} presents close-up views of the normalized out-of-plane current density $J_z$, electron velocity $U_{ey}$, and nongyrotropy measures $\sqrt{Q_e}$ and $D_\text{ng,e}$ downstream of the shock front, spanning $45\,d_{i,\text{up}}$ in x and $40\,d_{i,\text{up}}$ in y, at three consecutive snapshots from runs 14 and 15, which differ only in the upstream Mach number, overplotted with in-plane magnetic field lines.
We identify the reconnection X and O points by finding the local saddle/extrema from the 2D magnetic flux function $\psi$, defined such that $\mathbf{B} = \nabla \psi \times \hat{z}$ (i.e. $B_x = \partial \psi /\partial y$ and $B_y = -\partial \psi /\partial x$).

In both cases, we observe magnetic reconnection occurring downstream of the shock front. This process is marked by the formation of current sheets (first row in Figure \ref{fig8}), which subsequently break up to form magnetic X-points (reconnection sites) and O-points (plasmoids). Current sheets form intermittently downstream of the shock front, which can be directly linked to the steepening and compression of foreshock waves as they are convected through the shock transition.
During the shock reformation process at $t = 7.2\,\omega_{ci,\text{up}}^{-1}$ in Run 15, we find a bunch of small reconnection sites and plasmoids between the previous shock front and the newly forming shock front (Figure \ref{fig8}f).
Comparing Run 14 ($\text{M}_\text{A}=7$) and Run 15 ($\text{M}_\text{A}=12$) reveals a clear dependence on the Mach number. The higher Mach number in Run 15 leads to a larger out-of-plane current density $J_z$, more strongly curved magnetic field lines, and a significantly higher number of local reconnection X/O points. The plasmoids generated in the reconnection outflow have a length scale of up to $5\,d_{i,\text{up}}$.

An analysis of electron kinetic features, such as the nongyrotropy measure $\sqrt{Q_e}$ (\cite{swisdak2016quantifying}, third row in Figure \ref{fig8}) and $D_{ng,e}$ (\cite{aunai2013electron}, forth row in Figure \ref{fig8}), shows that these quantities are enhanced not only near reconnection separatrices but also at the main shock front itself. This indicates that in the turbulent shock environment, other kinetic processes—such as wave-particle interactions at the shock front and instabilities within the compressed plasma—also contribute significantly to the generation of nongyrotropy.

\begin{figure}
\noindent\includegraphics[width=0.9\textwidth]{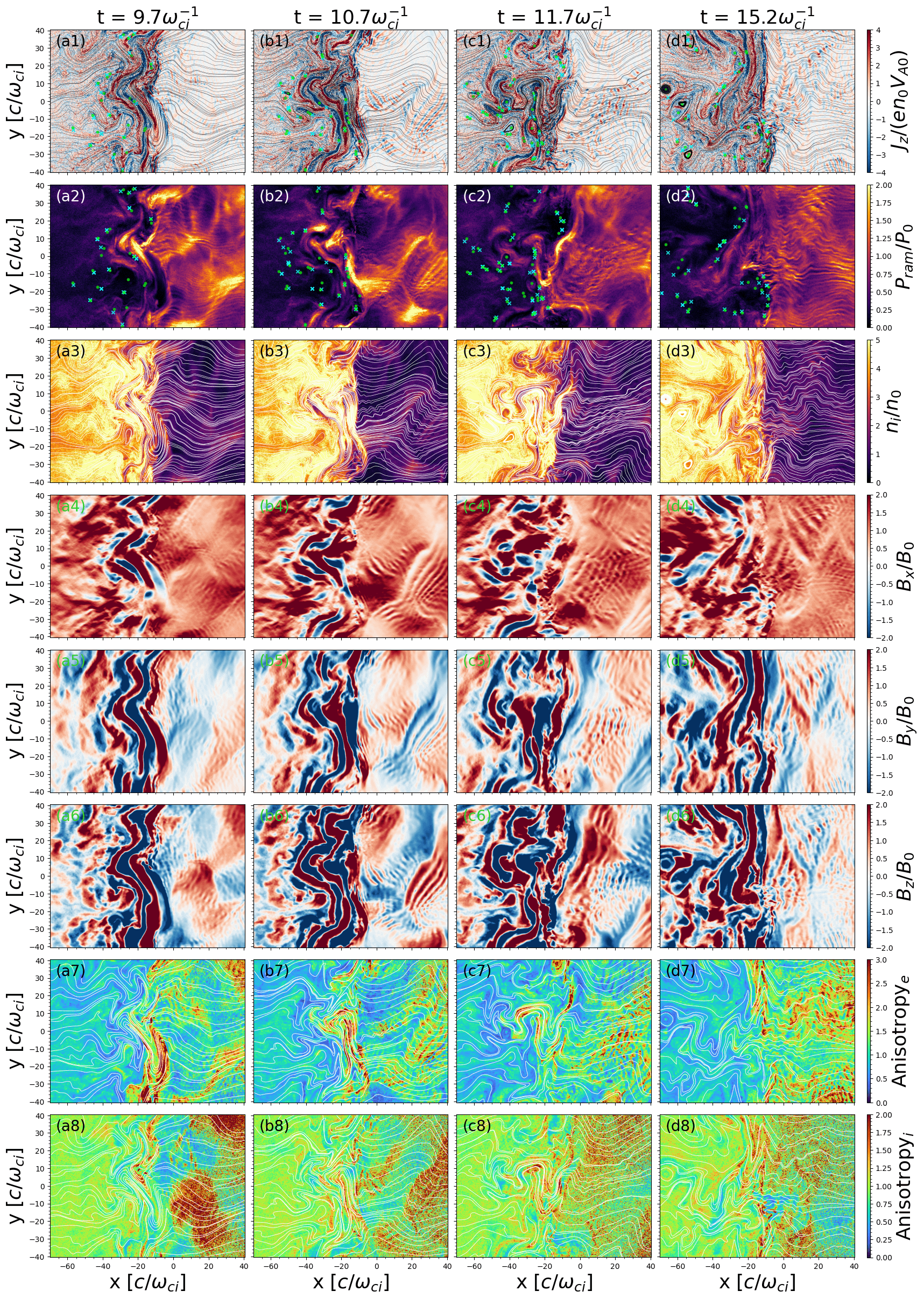}
\caption{Colored contours of out-of-plane current density $J_z$, ram pressure $P_\text{ram}$, ion density $n_i$, magnetic fields and electron and ion anisotropies in the 2D $\theta_{Bn}=0^\circ, \mathrm{M}_\mathrm{A}=7$ quasi-parallel shock run 16 at four consecutive snapshots, overplotted with in-plane magnetic field lines. Reconnection X-points and O-points are marked with cyan crosses and green circles, respectively.}
\label{fig9}
\end{figure}

To further explore these dynamics in a different geometry, we present various quantities from the $\theta_{Bn}=0^\circ$ parallel shock (run 16) in Figure \ref{fig9}.
We zoom-in to a local region that covers about $50\,d_{i,\text{up}}$ upstream/downstream from the shock front and $80\,d_{i,\text{up}}$ wide in the shock tangential direction.
Magnetic field lines are shown in black and white on top of the colored variables.
This run shares the same Alfvén Mach number as run 14. We find many more local reconnection X (cyan crosses) and O (green circles) points in the fully-parallel case, which are possibly linked to the more turbulent current sheet formation and reconnection processes in the shock transition layer for the small cone angle case. The out-of-plane current density $J_z$ also generates more curved magnetic fields.

The second row in Figure \ref{fig9} shows the ram pressure $P_\text{ram}$, highlighting jet structures propagating downstream across the shock front, which dissipates over time. The jets in this full PIC simulation can penetrate up to $20\,d_{i,\text{up}}$ into the downstream region before dissipation.
We also observe jets with bow-wave-like structures at the front (Figure \ref{fig9} a2-c2 near $x = -40\,d_{i,\text{up}}$).
In the upstream of the shock, we can see clear transverse magnetic oscillations from the magnetosonic-whistler modes (Figure \ref{fig9} row 5-6). At a later time $t = 15.2\,\omega_{ci,\text{up}}^{-1}$, filaments of ion density $n_i$ along the magnetic field direction appear with length scale $\sim 2\,d_{i,\text{up}}$ in the transverse direction.
These filaments are quasi-equilibrium structures after the saturation of Weibel instability \citep{morse1970} with an anti-correlation between the dynamic pressure and magnetic pressure.

\section{Discussion}

Through decades of local planar kinetic simulations, we have already accumulated decent understandings of micro-scale dynamics in collisionless shocks.
\texttt{FLEKS} was created with the goal of embedding local kinetic processes into a global MHD system, but it was not applied to study kinetic shock physics in the past due to the computational constraints and numerical difficulties.
Here we demonstrate that \texttt{FLEKS}, with an improved field solver, can reveal the key kinetic processes and ULF waves across planar shocks at different shock-normal angles, using the average solar wind conditions at 1 AU from WIND measurement.
We notice that for Earth-like bow shocks with $\beta \simeq 1$ the ion phase space holes are less obvious than the low-$\beta$ cases as shown by many previous local full-PIC shock simulations, $\beta_i \simeq \beta_e \simeq 0.1$ (e.g. \citet{umeda2006full, nakanotani2022collisional}).
This can be easily understood as the larger thermal velocity extent smears out the holes in the velocity space, and different structures tend to overlap with each other.
By experimenting with different initial shock states, we confirm that \texttt{FLEKS} also gives consistent results with previous local shock studies under different scenarios, for example the low Mach number ($\text{M}_\text{A} \sim 3$) subcritical quasi-perpendicular shock regime (MMS observation \citep{graham2025structure} and hybrid simulations \citep{ofman2009shock, ofman2013shock} with periodic downstream structures under pressure equilibrium), the low-$\beta$ supercritical shock regime \citep{scholer2003quasi} with the perpendicular shock reformation cycle, moderate Mach number ($\text{M}_\text{A} \sim 10$) supercritical parallel shock regime with downstream reconnection and jets \citep{wang2024origin, ren2024hybrid}, and high Mach number ($\text{M}_\text{A} \sim 60$) supercritical shocks with filamentary Weibel instability \citep{morse1970} and non-resonant streaming instability (i.e. Bell instability \citep{bell2004turbulent}).
The comparisons clearly show that collisionless shock behaviors are very sensitive to the specific plasma regimes.
These checks are useful for guiding the future global bow shock simulations under different upstream solar wind parameters, including but not limited to the average Mach number and plasma $\beta$ as we present above for validation, and also link back to the local studies of detailed shock physics.

\subsection{Dimensionality: 1D Artifacts and the Need for Higher Dimensions}

Early attempts to simulate mirror and EMIC waves often relied on simplified 1D hybrid simulations initialized with bi-Maxwellian ion distributions exhibiting high anisotropy ($P_{i\perp} / P_{i\parallel} = 6$), as exemplified by \citet{price1986numerical}.
These simulations, however, lacked a self-consistent treatment of the shock and its role in generating the anisotropy.
Our experiments demonstrate that when the shock is included, the constraints imposed by reduced dimensionality become crucial.
It is important to recognize that our 1D simulation (Figure \ref{fig1}) neglects parallel heating along the z-direction.
Consequently, no proton thermalization occurs along $V_{iz}$ in the downstream and ramp regions (Figure \ref{fig1}j and \ref{fig1}n).
This leads to highly anisotropic downstream ion distributions ($P_{i\perp}/P_{i\parallel} > 8$), a clear artifact of the reduced dimensionality.
As demonstrated in the 2D quasi-perpendicular simulation (Figure \ref{fig3}), downstream parallel and oblique propagating waves, absent in the 1D case, play a crucial role in redistributing free energy associated with temperature anisotropy.
This finding demonstrates that 1D simulations are insufficient for capturing the downstream waves generated by perpendicular shocks.
The 2D simulation, incorporating realistic solar wind conditions upstream of Earth \citep{salem2023precision}, provides a more accurate representation of the terrestrial bow shock compared to the 1D case, underscoring the importance of at least 2D modeling for capturing the complex dynamics of quasi-perpendicular shocks, including the interplay of anisotropy-driven instabilities and wave generation.

It is also crucial to acknowledge certain limitations inherent in the 1D planar shock setup for quasi-parallel shocks.
Firstly, with only one spatial degree of freedom, wave propagation is constrained to the x-direction.
We established our background upstream magnetic field at a shock-normal angle $\theta_{Bn} = 30^\circ$ relative to the -x direction.
Consequently, this oblique wave angle remains fixed and may not align with the maximum wave growth direction predicted by linear theory.
However, \citet{lucek2002parallel} reported from in-situ satellite measurements that the average $\theta_{Bn}$ is approximately $20^\circ$, a value inconsistent with the prediction for maximum parallel propagating linear wave growth.
This discrepancy illustrates the complexity of wave behavior in real-world environments compared to idealized simulations.
Secondly, the periodic condition applied in the y and z direction limits the growth of structures that are intrinsically high-dimensional, such as downstream magnetic reconnection, plasma jets and return flows, which are observed in our 2D simulations (Figure \ref{fig7}, \ref{fig8} and \ref{fig9}).

\subsection{Interpreting Downstream Structures in 1D}

In our 1D quasi-perpendicular simulation, correlated periodic downstream fluctuations are observed (Figure \ref{fig1}a, c, d). These postshock oscillations are not indicative of transmitted or generated waves. Instead, they are periodic structures that arise from ion ring distributions. This is true despite their exhibiting density and magnetic field correlations similar to compressional fast magnetosonic waves \citep{russel2009stereo}.
Interpretations of these structures as damping solitons \citep{sagdeev1966shock, kennel1967shock} are inconsistent with the shock's $\beta$ and Mach number.
These structures also differ from the stationary downstream oscillations analyzed in \citet{ofman2009shock, ofman2013shock}, where subcritical shocks maintain total pressure balance with anti-correlated magnetic field and ion pressure variations.
While their descriptions involve non-gyrotropic downstream ions, our simulations clearly show a symmetric velocity distribution in the perpendicular $V_{ix}-V_{iy}$ plane.
The ramp width in our simulation aligns with observations from early AMPTE satellite measurements \citep{walker1999thin, walker1999ramp} and subsequent Cluster and THEMIS data \citep{hobara2010stats}. Furthermore, the revealed ion reflection is consistent with the model proposed in the Appendix of \citet{khotyaintsev2024ion}, which falls into the category where ions, upon reflection, are decelerated by $\Delta \phi$ and then undergo cyclotron gyration due to the increased magnetic field in the ramp, leading to a broad distribution of the reflected population within the ramp.
The inclusion of electron kinetics in \texttt{FLEKS} introduces electrostatic electron pressure modulation by ion and magnetic pressure variations (Figure \ref{fig1}c, h). This effect is absent in hybrid models that treat electrons as a massless fluid in an ideal thermodynamic process (e.g. adiabatic if $\gamma=5/3$; isothermal if $\gamma=1$).

\subsection{Foreshock Wave Dynamics: Shocklets, SLAMS, and Whistlers}

Our simulations of quasi-parallel shocks capture the nonlinear evolution of upstream ULF waves. As the fast magnetosonic waves are convected back towards the shock by the super-Alfvénic upstream flow, they evolve into \textit{shocklets} \citep{hoppe1981upstream, wilson2016low}, i.e. steepened flanks characterized by small-scale fluctuations ($\delta B / B_0 < 2$) that function as miniature shocks.
\textit{Short, Large Amplitude Magnetic Structures} (SLAMS, or magnetic pulsations, \citet{schwartz1992observations}) share similarities with shocklets but are defined by larger magnetic field fluctuations ($\delta B / B_0 > 2$) and potential soliton-like behavior.
Consequently, our nonlinearly steepened ULF wave signatures (Figure \ref{fig2}) align more closely with observed shocklets, where nonlinear evolution is driven by interactions with energetic particle pressure gradients, and the ensemble of these structures forms the shock transition (e.g., \citep{dubouloz1995two}).
The downstream magnetic fluctuations also exhibit higher amplitudes and broader spatial extents, consistent with observational data from Earth's quasi-parallel magnetosheath (e.g., \citet{lucek2002parallel}).

The presence of whistler precursors is sensitive to the mass ratio (Figure \ref{fig-B-compare}). We observe clear whistler waves upstream of the shock front with wavelength $\lambda_w \sim 2d_{i,\text{up}}$ in larger mass ratio cases ($m_i/m_e = 100, 400$), where the whistler critical Mach number $\text{M}_\text{w}\gtrsim \text{M}_\text{A}$, consistent with the formation criterion (Equation \ref{eq-whistler-mach}).
These whistler waves consistently appear on the leading edges in the first few ULF wavelengths ahead of the shock, which may be linked directly to the main shock front or the upstream steepened shocklets \citep{wilson2016low}.
Similar whistler precursor occurrences are reported in full PIC simulations, e.g. \citet{tsubouchi2004slams, nakanotani2022collisional}.

\subsection{Downstream Jets and Reconnection}

Our 2D quasi-parallel simulations show localized enhancements in ram pressure characteristic of jets \citep{plaschke2018jets}, which are frequently observed in the terrestrial magnetosheath. In run 14 (Figure \ref{fig7}), we find an occurrence rate of approximately one jet per $2\pi/\omega_{ci,\text{up}}$, or roughly 4 jets per minute.
Despite the lack of a clear definition for jets, our simulations reveal a jet occurrence rate higher than observational estimates of a few jets per hour.
This discrepancy might be partly attributed to the generally smaller scale of our simulated jets ($\lesssim 10\,d_{i,\text{up}}$) compared to observed terrestrial magnetosheath jets, making them more challenging for sparse satellites to detect.
The observed return flow (Figure \ref{fig7}b) is similar to satellite observation of sunward flows in the downstream magnetosheath (e.g., \citet{archer2014role}).

Furthermore, our simulations show that these jets dissipate rapidly. In the parallel shock case (Figure \ref{fig9}), the jets penetrate up to $20\,d_{i,\text{up}}$ before dissipation. We also observe bow-wave-like structures at the jet front (Figure \ref{fig9} a2-c2), consistent with the hybrid simulations of \cite{ren2024hybrid}.
However, the jet scale sizes are shorter than the $\sim 100\,d_{i,\text{up}}$ penetration depth reported in \cite{ren2024hybrid}. Part of this difference arises from the inclusion of kinetic physics across the electron and ion scales in the FLEKS model. Subion-scale micro-instabilities, which are absent in hybrid simulations where electrons are treated as a fluid, may be excited by the strong pressure anisotropies (Figure \ref{fig9} row 7-8) and inhomogeneous flows near the shock front.
%In the saturation phase, the kinetic energy of the streaming particles has been converted into thermal energy and magnetic energy, with a quasi-static equilibrium being established and an anti-correlation between the ram pressure and magnetic field.
These instabilities could provide an additional dissipation channel that hinders the deep penetration of jets, indicating a key physical process that requires a full kinetic treatment to be captured. This limited penetration in our 2D planar shock simulations, along with the inherent constraint of the two-dimensional setup, offers plausible explanations for the observed model-observation mismatch in jet occurrence and scale.
Finally, our analysis of X/O points and electron nongyrotropy measures (Figure \ref{fig8} row 3-4) confirms the presence of shocked reconnection \citep{wang2024origin} but also shows that in this turbulent environment, other kinetic processes also contribute significantly to nongyrotropy, making it a challenging metric for uniquely identifying reconnection sites.

\subsection{Electron-to-Ion Temperature Ratio}

In-situ measurements of Earth's and Saturn's bow shocks (summarized in \citet{raymond2023electron}, Figure 1) reveal that the downstream electron-to-ion temperature ratio $T_e / T_i$ decreases from 1 at low Alfvén Mach numbers $\text{M}_\text{A}$ to around 0.1-0.3 at $\text{M}_\text{A}\sim 10$.
This trend is qualitatively consistent with a statistical study of interplanetary shocks by \citep{wilson2020electron}, which showed a large scatter but an average decrease in $T_e / T_i$ from 1 at low Mach numbers to $\sim 0.05$ at $\text{M}_\text{A}\sim 10$, with little dependence on upstream $\beta$ or shock obliquity $\theta_{Bn}$.
For the $\text{M}_\text{A} = 7$ shocks in our simulations, $T_e / T_i$ falls within the range of 0.2-0.4, aligning with predictions from several perpendicular shock simulations summarized in \citet{raymond2023electron} (Figure 2).
For quasi-perpendicular shocks, we find comparable $T_e / T_i \simeq 0.28$ in 1D (Figure \ref{fig1}c) and $T_e / T_i \simeq 0.26$ in 2D with in-plane magnetic field geometry (Figure \ref{fig3}f).
In our quasi-parallel shock simulations, while both electron and ion temperatures increase, the $T_e / T_i$ ratios (0.42 in 1D and 0.33 in 2D) are higher than in the corresponding quasi-perpendicular cases. This indicates preferential heating in quasi-parallel shocks and hints that 2D simulations predict higher $T_e / T_i$ ratios compared to 1D simulations.

\subsection{Numerical Performance and Model Limitations}

Our grid resolution study (Figure \ref{fig-res-compare}) highlights the stability of the semi-implicit PIC algorithm. The coarsest resolution $\Delta x = 4\,d_{e,\text{up}}\simeq 1\,d_{i,\text{up}}$ $(m_i / m_e = 25)$ is close to the ion inertial length.
Explicit full-PIC methods generally cannot work at this resolution due to severe finite-grid instability and numerical heating \citep{birdsall2018plasma}.
In contrast, \texttt{FLEKS} produces stable solutions that stay meaningful at sub-ion scales, demonstrating the algorithm's robustness in overcoming the constraints of explicit PIC methods.

A remaining consideration is the presence of an upstream-moving diffuse ion population extending to the inflow boundary in the quasi-parallel runs (Figure \ref{fig4}s-t, Figure \ref{fig6}).
While they energize waves and may be scattered back downstream, our simulations consistently show some ions reaching the boundary.
To maintain stability, we enforce a fixed Maxwellian distribution in the boundary ghost cells, effectively removing these ions.
This artificial boundary condition, while not causing significant issues in our local planar shock simulations, represents an important consideration for future work, particularly in the context of global MHD-AEPIC coupled simulations.
Investigating the transmission of waves and particles through the MHD-AEPIC boundary and its influence on the coupled system will be crucial for accurately capturing the global dynamics of the magnetosphere.
For a detailed discussion of the current MHD-AEPIC coupling scheme in our model and related fast and whistler wave tests, interested readers are referred to \citet{daldorff2014two, shou2021magnetohydrodynamic, chen2023fleks}.

\section{Conclusion and Future Work}

This study presents new applications of the improved semi-implicit energy-conserving particle-in-cell (PIC) model \texttt{FLEKS}, focusing on collisionless shock simulations.
The refined semi-implicit energy-conserving algorithm in \texttt{FLEKS} proved robust and efficient, enabling stable and accurate shock simulations with grid resolutions at sub-ion scales.
We conducted systematic one- and two-dimensional numerical experiments to validate the model under typical hypersonic, $\beta \sim 1$ shocks observed at the terrestrial magnetosphere and other astrophysical systems.
Our simulations successfully captured the characteristic features of both quasi-perpendicular and quasi-parallel shocks, including shock structures, wave generation, particle dynamics and downstream magnetic reconnection.
Importantly, we highlighted the necessity of at least two spatial dimensions for accurately capturing the full physics of collisionless shocks.

The results of these parameter studies provide valuable guidance for selecting physical and numerical parameters in global supersonic, super-Alfvénic magnetosphere simulations.
Given current computational constraints, performing three-dimensional global shock simulations with realistic mass ratios and fully resolved electron physics remains challenging.
However, using a reduced mass ratio, even at $m_i / m_e = 25$, can still yield meaningful insights into kinetic processes as revealed from the semi-implicit PIC local shock simulations.
We anticipate that global magnetosphere simulations incorporating full ion kinetics and appropriately modified electron physics will be instrumental in bridging the gap between large-scale MHD phenomena and small-scale kinetic processes.

Several avenues for future research emerge from this work.
Regarding MHD-AEPIC coupling, recent work by \citet{toth2024weak} proposed an extended MHD anisotropic pressure solver to control the partitioning of non-adiabatic heating between electrons and ions across discontinuities and shocks.
This development is potentially crucial for global MHD-AEPIC coupling, particularly in scenarios where the PIC domain covers only a portion of the shock, as it could mitigate numerical artifacts arising from inconsistent treatment of pressures and heat fluxes in the fluid and kinetic equations.
Implementing this new pressure solver within the Space Weather Modeling Framework for coupled MHD-AEPIC simulations is a promising direction for future investigation.

Furthermore, to address more realistic bow shock geometries, large-scale 2D/3D coupled global magnetosphere simulations are required to investigate the dynamics of the curved shocks and downstream jets.
This represents a crucial step beyond planar shock simulations, which can further capture the spatial separation of electron and ion foreshock regions due to the difference of electron and ion masses and time-of-flight effects.
Global simulations will enable a more comprehensive understanding of the interplay between shock geometry, particle dynamics, and wave generation in the planetary magnetospheres.

%% Please use the acknowledgment and contribution environments. This will 
%% be anonomyized when the "anonymous" style option is used. 
\begin{acknowledgments}
This work was supported by NASA grants 80NSSC23K1409 and 80NSSC24K0144, DOE grant DE-SC0024639 and the Alfred P. Sloan Research Fellowship.
We thank Dr. Adnane Osmane, Dr. Yufei Hao, Dr. Xin An, and Dr. Zhongwei Yang for the insightful discussions on the shock wave properties.
We also thank the reviewer for the constructive comments that greatly improve the manuscript.
We would like to acknowledge high-performance computing support from the NASA High-End Computing Program through the NASA Advanced Super-computing Division at Ames Research Center, from National Energy Research Scientific Computing Center, a DOE Office of Science user facility, and from the Derecho system (\texttt{doi:10.5065/qx9a-pg09}) provided by the NSF National Center for Atmospheric Research (NCAR), sponsored by the National Science Foundation.  
\end{acknowledgments}

\begin{contribution}

HZ came up with the initial research concept and was responsible for designing the tests, validating the results, writing and submitting the manuscript.
YC obtained the funding and designed the updated algorithm.
CD supervised the project and provided ideas for application in global studies.
LW conducted comparative simulations to validate the results.
YZ also provided the funding and ideas about downstream wave studies.
BW provided feedbacks on the modeling results and raised interesting discussions about energy transport.
GT offered suggestions on algorithm improvements and insights in interpreting the scalings.
All authors reviewed and contributed to the manuscript. 

\end{contribution}

%\facilities{HST(STIS), Swift(XRT and UVOT), AAVSO, CTIO:1.3m, CTIO:1.5m, CXO}

\software{SWMF \url{https://github.com/SWMFsoftware/SWMF},  
          FLEKS \url{https://github.com/SWMFsoftware/FLEKS}, 
          Batsrus.jl \citep{batsrusjl},
          fleskpy \citep{flekspy},
          }


%\appendix

%\section{Appendix information}

%% For this sample we use BibTeX plus aasjournalv7.bst to generate the

\bibliography{ref}{}
\bibliographystyle{aasjournalv7}


% \listofchanges

\end{document}